\documentclass[10pt,journal]{IEEEtran} 

\usepackage[utf8]{inputenc}
\usepackage[T1]{fontenc}

\usepackage{amssymb,amsmath}
\usepackage{cite}
\usepackage{psfrag}
\usepackage{url}
\usepackage[absolute,overlay]{textpos}

\DeclareMathOperator{\Tr}{Tr}
\usepackage{pgf}
\usepackage[export]{adjustbox}
\usepackage{color}
\usepackage{bm}
\usepackage{bbm}
\usepackage{booktabs}
\usepackage{url}
\usepackage{upgreek}
\usepackage{verbatimbox}
\usepackage{graphicx}
\usepackage{stackengine}

\usepackage{algorithm}
\usepackage{algorithmic}

\usepackage{amsthm}
\usepackage{afterpage}
\usepackage{enumitem,lipsum}

\usepackage{setspace}
\begin{document}
	
	\title{Massive MIMO with Radio Stripes for Indoor Wireless Energy Transfer}
	\author{Onel L. A. López,~\IEEEmembership{Member,~IEEE,}
		Dileep Kumar,~\IEEEmembership{Graduate Student Member,~IEEE,}
		Richard Demo Souza,~\IEEEmembership{Senior Member,~IEEE,}
		Petar Popovski,~\IEEEmembership{Fellow,~IEEE,}
		Antti T\"olli,~\IEEEmembership{Senior Member,~IEEE,}
		and Matti Latva-aho,~\IEEEmembership{Senior Member,~IEEE}
		\thanks{Onel López, Dileep Kumar, Antti T\"olli and Matti Latva-aho are  with  the Centre for Wireless Communications University of Oulu, Finland, e-mails: \{Onel.AlcarazLopez, Dileep.Kumar, Antti.Tolli, Matti.Latva-aho\}@oulu.fi.}
		\thanks{Richard Demo Souza is with Federal University of Santa Catarina (UFSC), Florianópolis, Brazil, e-mail: richard.demo@ufsc.br.}
		\thanks{Petar Popovski is with the Department of Electronic Systems, Aalborg University, 9220 Aalborg, Denmark, e-mail: petarp@es.aau.dk}
		\thanks{This work is supported by Academy of Finland (Aka) (Grants n.307492, n.319059, n.318927 (6Genesis Flagship)), CNPq, Print CAPES-UFSC ``Automation 4.0'', and RNP/MCTIC (Grant  01245.010604/2020-14) 6G Mobile Communications Systems. The work of Petar Popovski has been partially supported by EU H2020 RISE-6G project.}
	} 
	
	\maketitle
	\begin{abstract}		
		Radio frequency wireless energy transfer (WET) is a promising solution for powering autonomous Internet of Things (IoT) deployments. In this work, we leverage energy beamforming  for  powering multiple user equipments (UEs) with stringent energy harvesting (EH) demands in an indoor distributed massive multiple-input multiple-output system. Based on semi-definite programming, successive convex approximation (SCA), and maximum ratio transmission (MRT) techniques, we derive optimal  and sub-optimal precoders aimed at minimizing the  radio stripes' transmit power while exploiting  information of the power transfer efficiency of the EH circuits at the UEs. Moreover, we propose an  analytical framework to assess and control the electromagnetic field (EMF)  radiation exposure in the considered indoor scenario. Numerical results show that i) the EMF radiation exposure can be more easily controlled at higher frequencies at the cost of a higher transmit power consumption, ii) training is not a very critical  factor for the considered indoor system, iii) MRT/SCA-based precoders are particularly appealing when serving a small number of UEs, thus, especially suitable for implementation in a time domain multiple access (TDMA) scheduling framework, and iv) TDMA is more efficient than spatial domain multiple access (SDMA) when serving a relatively small number of UEs. Results suggest that additional boosting performance strategies are needed to increase the overall system efficiency, thus making the technology viable in practice. 
	\end{abstract}
	\section{Introduction}\label{intro} 	
	The expected massive number of ``Internet of Things'' (IoT) devices coming online over the next decade is conditioned on first solving critical challenges, especially in terms of efficient massive access, lightweight communication protocols, and sustainable powering mechanisms. As for the latter, existing solutions that rely on wired powering are usually cost-prohibitive or infeasible for ubiquitous deployment. Meanwhile battery-powered solutions face other drawbacks \cite{LopezAlves.2021}: i) limited lifetime, influenced by activity/usage, ii) frequent maintenance as different devices, and corresponding lifetimes, may coexist in the same environment, and iii) the battery waste processing problem. Therefore, alternative powering mechanisms and related technologies need to be developed in the coming years to realize the vision of a data-driven sustainable society, enabled by near-instant, secure, unlimited and green massive connectivity \cite{LopezAlves.2021,Mahmood.2020}.	
	
	Radio frequency (RF) wireless energy transfer (WET) constitutes an appealing technology to be further researched, developed and exploited for powering IoT deployments  \cite{LopezAlves.2021,Mahmood.2020,Huang.2015,Clerckx.2021}. Different from near-field WET solutions as those exploiting induction, magnetic resonance coupling and piezoelectricity phenomena, and 	
	energy harvesting (EH) 
	from other energy sources, RF-WET allows: i) small-form factor, ii) native multi-user support, and iii) relatively long range energy coverage.\footnote{Note that methods relying on near-field WET or EH from, e.g., light intensity, 	thermal energy or even wind, are either highly sensitive to blocking or have low conversion efficiency. But maybe more importantly, they demand an add-on EH material and circuit, which in practice limits the form factor reduction to the desired levels for many use cases \cite{LopezAlves.2021}.} Observe that independently of the energy source, WET and EH processes occur fundamentally at different nodes: WET is a process initiated at an energy transmitting node, while EH occurs at an energy receiving node. Meanwhile, in ambient EH, the energy transmitters are not intentionally powering surrounding  devices, which instead perform opportunistic EH. However, ambient RF-EH (at least as a standalone) does not suffice for most use cases of practical interest due to the very limited and uncontrolled energy transfer. Instead (or as a complement), dedicated RF-WET (hereinafter just referred to as WET) is usually required.

	Unfortunately, the ultra-low end-to-end power transfer efficiency (PTE) of WET systems, together with strict regulations on the electromagnetic field (EMF) radiation, are critical factors that may limit WET feasibility in practice. By addressing these challenges, while exploiting further advancements on ultra-low-power integrated circuits, WET may emerge as a revolutionary technology that will finally cut the \textit{``last wires''} (i.e.,  cables for energy recharging \cite{Huang.2015}) for a truly wireless and autonomous connectivity. In fact, IoT industry is already strongly betting on this technology, proof of which is the variety of emerging enterprises with a large
	portfolio of WET solutions, e.g., Powercast, TransferFi and Ossia\footnote{See https://www.powercastco.com, https://www.transferﬁ.com and https://www.ossia.com.}.  
	
\begin{figure}[t!]
		\centering
		\includegraphics[width=0.46\textwidth]{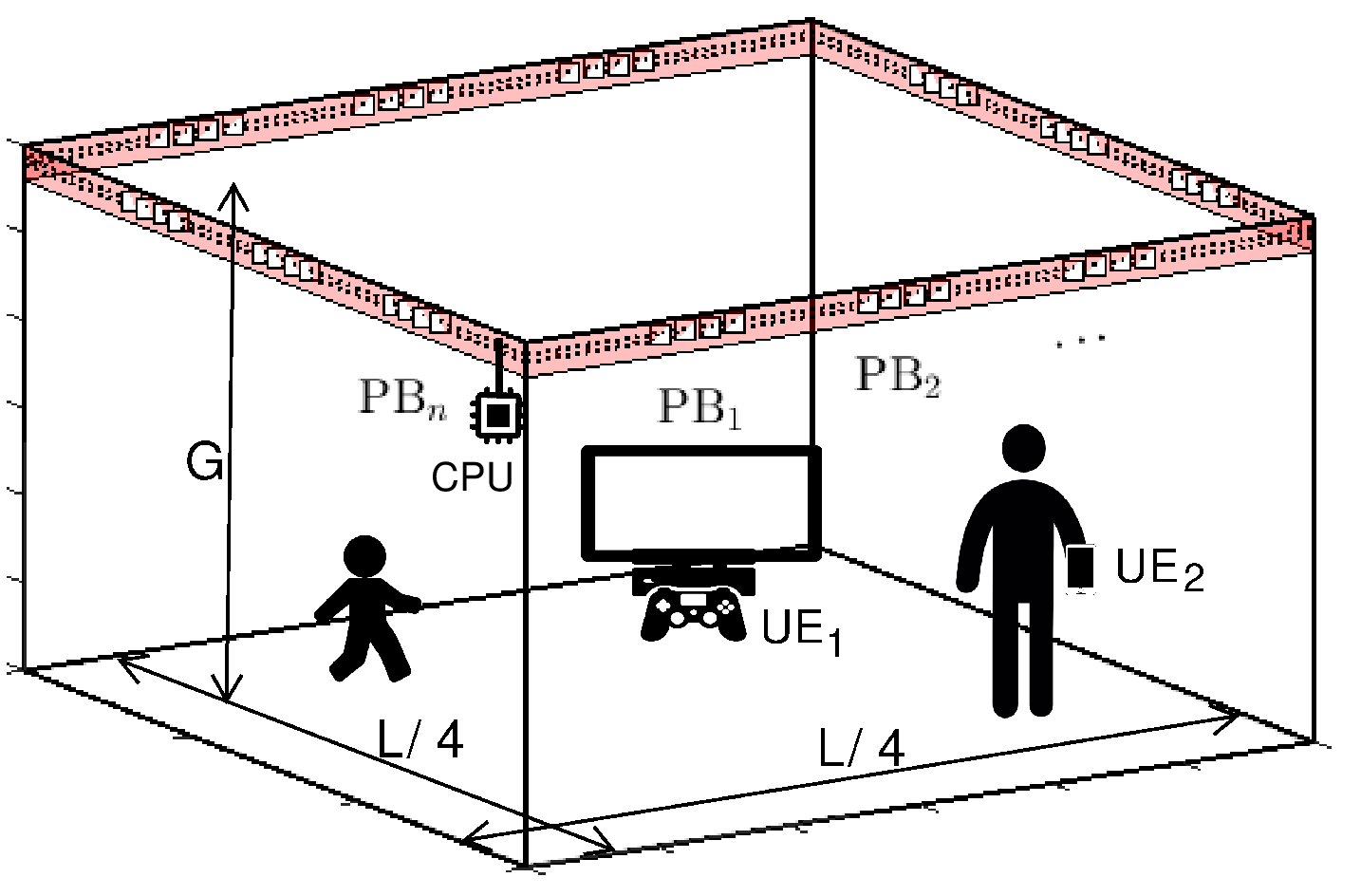}\\
		\includegraphics[width=0.46\textwidth]{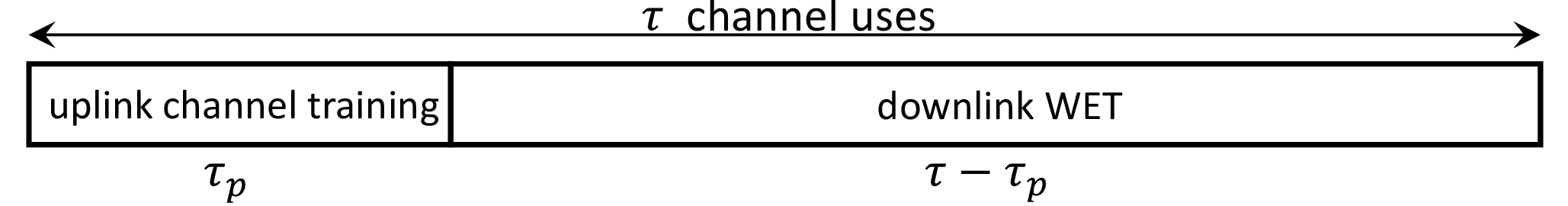}
		\caption{System model with $K=2$, $N=12$ and $M=4$ (top), and structure of the training/charging protocol (bottom).}
		\label{Fig1}
	\end{figure}

	Recent works on WET have mainly targeted  extremely low-power/cost IoT applications, e.g., wireless sensor networks and RFIDs, due to the ultra-low  	PTE (see \cite{LopezAlves.2021} and references therein). However, several  promising technologies such as i) energy beamforming and waveform optimization \cite{Clerckx.2021,Clerckx.2016,LopezMontejo.2020,Lopez.2021}, ii) distributed \cite{Rosabal.2020,Van.2020} and massive antenna \cite{Yang.2015,Khan.2018,Lee.2018,Wang.2021,Zhu.2019,ZhaoLong.2019,Van.2020} systems, iii) smart reflect arrays and reconfigurable metasurfaces \cite{Pan.2020}, iv) motor-equipped power beacons (PBs) \cite{Zhang.2018}, flying  PBs \cite{XuZeng.2018}, PBs with rotary antennas \cite{Lopez.2021}), and v) mobile computation offloading and crowd sensing \cite{Clerckx.2021}, may broaden WET applicability, and turn it plausible for powering more energy-hungry IoT devices, e.g., smartphones, game console controllers, electronic toys. 
	In this work, we leverage some of above enablers to investigate WET's feasibility for power-hungry indoor charging. Specifically, optimal and sub-optimal energy beamforming schemes complying with EMF-related constraints are proposed and discussed for a distributed massive multiple-input multiple-output (mMIMO) radio stripes system deployed in an indoor small-area environment (thus with limited serving distances) for multi-user RF wireless charging as illustrated in Fig.~\ref{Fig1}. 
	\subsection{Related Work}\label{RW}
	The max-min throughput optimization problem for a mMIMO wireless powered communication network (WPCN) is addressed in \cite{Yang.2015}. Therein, authors show that the asymptotically, in the number of transmit antennas, optimal energy beamformer is akin to maximum ratio transmission (MRT) in MIMO communications. Taking advantage of the latter results, authors in \cite{Khan.2018} investigate the  overall PTE and energy efficiency of a WPCN, where a mMIMO base station uses WET to charge single-antenna EH users in the downlink. A piece-wise linear EH model just considering sensitivity and saturation impairments is adopted, and it is shown that increasing the transmit power improves the energy efficiency as the number of antennas becomes large. However,  authors in \cite{Yang.2015,Khan.2018} assume that all the harvested energy is used for uplink information transmission, ignoring other important energy consumption sources. Meanwhile, a low-complexity mMIMO WET scheme based on the retrodirective beamforming technique, where all EH devices send a common beacon  simultaneously to the  PB in the uplink and the PB simply conjugates and amplifies its received sum-signal for downlink WET, is investigated in \cite{Lee.2018} relying on the asymptotically optimum MRT-like precoding~\cite{Yang.2015}.  The performance in terms of received net energy (harvested energy minus energy consumed during training) of a multi-user mMIMO system with Rician fading channels is analyzed in \cite{Wang.2021}. Two different schemes are considered: i) training-based WET, and ii) line-of-sight (LOS) beamforming, and authors derive a  user-specific path loss threshold for 
	switching between them. The appeal of LOS-based beamforming is also illustrated in \cite{Lopez.2021}, where an MRT-based energy beamforming algorithm with low complexity for mMIMO is proposed. However, the system performance is analyzed only in the ultra-low EH regime in \cite{Wang.2021,Lopez.2021}, while the channel assumption in \cite{Wang.2021} is also very restrictive. Finally, readers may refer to \cite{Zhu.2019,ZhaoLong.2019} where authors investigate the performance of mMIMO-enabled simultaneous wireless information and power transfer systems.
	
	As evinced above, single-transmitter mMIMO technology for WET has been considerably investigated. Surprisingly, much less efforts have been put to investigate distributed mMIMO technology, which makes WET more feasible by shortening the channel distances \cite{Van.2020}. To the best of our knowledge, there is no prior work addressing mMIMO with radio stripes \cite{Frenger.2019,Interdonato.2019} for WET. Note that in a radio stripes implementation, actual PBs consist of antenna elements and circuit-mounted chips inside the protective casing of a cable/stripe, thus, allowing  imperceptible installation and alleviating the problem of deployment permissions \cite{LopezAlves.2021}. Moreover, its distributed nature and the significantly reduced transmit power per PB lead to implementations with low heat dissipation and low-cost hardware. These features may make radio stripes technology appealing for supporting high-efficiency WET. We show later that it is indeed possible to deliver significant amounts of energy  over practical third generation partnership project (3GPP) channels to multiple EH receivers  by using this technology in an indoor small-area environment. 
	
	On the other hand, WET systems are subject to strict transmit constraints that limit their performance in practice. Specifically, the EMF caused by energy radiation must be strictly limited in the presence of humans (and other living species) to reduce the risks of potential biological effects (e.g., tissue heating) \cite{Ahlbom.1998,Tran.2017}. Concerns about adverse consequences of EMF  exposure have resulted in the establishment of exposure limits\footnote{EMF limits are set by the International Commission on Non-Ionizing Radiation Protection (ICNIRP) in most of Europe \cite{Ahlbom.1998}.} such as the maximum permissible exposure, also called power density, measured in $[\text{W}/\text{m}^2]$. Although several works have considered such important constraints in WET systems, e.g., \cite{Dai.2017,DaiZhao.2018}, to the best of our knowledge none has yet addressed them  in the context of radio stripes -enabled mMIMO WET.
	\subsection{Contributions} 
	This article considers an indoor mMIMO system with radio stripes for multi-user WET under EMF-related constraints. Our main contributions are:
	\begin{enumerate}[wide, labelwidth=!, labelindent=0pt]	
		\item We introduce an indoor mMIMO system with radio stripes for wirelessly charging user equipments (UEs). The radio stripes system is composed of multi-antenna PBs subject to maximum transmit power constraints, which coherently beamform the energy signals towards the devices using channel state information (CSI) obtained from uplink training, and PTE information of the EH circuits at the UEs. State-of-the-art precoders aiming at minimizing the radio stripes' transmit power and subject to stringent EH requirements per UE are adopted.
		\item We propose an analytical framework to evaluate the RF power density in the proximity of the UEs, and that caused at a random point in the network. We formulate these figures in terms of EMF constraints and incorporate them to the precoders' design accordingly. 
		\item We show that the spatial transmit gains from incorporating more PBs to the radio stripes at high frequencies do not compensate the path loss, demanding more transmit energy resources. However, the EMF radiation exposure can be more easily controlled at higher frequencies. 
		\begin{table*}[!t]
			
				\caption{Acronyms and Main Symbols}
				\label{tab1}
				\centering
				\small
				\begin{tabular}{p{0.95cm} p{7.3cm}p{0.95cm} p{7.5cm}}
					\toprule
					3GPP & third generation partnership project & CDF & cumulative distribution function\\ 
					CPU & central processing unit &  CSI & channel estate information\\
					DC & direct current & EH & energy harvesting \\
					EMF  & electromagnetic field & ICNIRP & Int. Commission on Non-Ionizing Radiation Protection\\
					IoT & Internet of Things & LOS & line-of-sight \\
					LS & least square & MIMO & multiple-input multiple output \\
					mMIMO &  massive MIMO & MRT & maximum ratio transmission \\
					NLOS & non-LOS & PB & power beacon \\
					PTE & power transfer efficiency & 	RF & radio frequency\\
					SAR & specific absorption rate & SCA & successive convex approximation\\
					SDMA & spatial division multiple access & SDP & semi-definite programming\\
					TDMA & time division multiple access & UE & user equipment \\
					WET & wireless energy transfer & WPCN & wireless powered communication network \\	
					\hline
					$\mathcal{N}$, $N$ & set, and number, of PBs in the radio stripes & $\text{PB}_n$,$\text{UE}_k$& $n-$th PB, $k-$th UE\\ 
					$M$ & number of antennas per PB & $K$ & number of single-antenna EH UEs \\
					$L$ & length (m) of the radio stripes & $G$ & height (m) of the room/office \\ $f,\lambda$ & operation frequency, associated wavelength & $\mathbf{h}_{kn}$ &
					fading coefficient of channel $\text{PB}_n\leftrightarrow\text{UE}_k$ \\ $\bar{\mathbf{h}}_{kn}$ & first order statistics (mean) of $\mathbf{h}_{kn}$&
					$\mathbf{R}_{kn}$ & covariance matrix of the realizations of  $\mathbf{h}_{kn}$ \\
					$\tau$ & number of channel uses in a coherence block & $\tau_p$ & number of pilot symbols\\
					$\bm{\psi}_i$ & $i-$th pilot ($\tau_p-$dimensional) vector signal & $p_k$ & transmit power of $\text{UE}_k$ \\
					$\mathbf{W}_n, \sigma^2$ & noise matrix, and associated power, at $\text{PB}_n$ &
					$\mathbf{Y}_n$ & receive pilot signal (matrix) at $\text{PB}_n$ \\ 		     $\hat{\mathbf{h}}_{kn}, \tilde{\mathbf{h}}_{kn}$ & LS estimate, and error, of $\mathbf{h}_{kn}$ & $K'$ & number of transmitted energy symbols \\
					$s_{k'}$ & $k'-$th energy symbol &  $\mathbf{v}_{k'n}$ & precoding vector associated to $s_{k'}$ \\
					$y_k$ & receive energy signal at $\text{UE}_k$ & $P_k$ & incident RF power per channel use at $\text{UE}_k$ \\
					$g_k(\cdot)$ & PTE/EH function  & $E_k$ & energy harvested by $\text{UE}_k$ in a block \\
					$\varpi_k,\nu_k$ & DC sensitivity, saturation level of EH at $\text{UE}_k$ & $\hat{E}_k,\hat{P}_k$ & estimated $E_k$, $P_k$ \\
					$\xi_k$ & target energy requirement per block at $\text{UE}_k$ & $\delta_k$ & RF energy requirement at $\text{UE}_k$ in WET phase\\
					$P_T$ & total transmit power of the radio stripes & $p_\text{max}$ & per-PB transmit power constraint\\
					$\epsilon$ & solution accuracy of optimization algorithm & $\widetilde{\hat{P}}_k$ & first-order Taylor approximation of $\hat{P}_k$ \\
					$\overline{\mathbf{v}}_{k'},\overline{\delta}_{k'n}$ & a fixed operating point of $\mathbf{v}_{k'}$, $\delta_{k'n}$ & $p_{kn}$ & transmit power budget of $\text{PB}_n$ to $s_k$\\  
					$\mathbf{q}_{k'n}$ & normalized MRT precoder of $\text{PB}_n$ for $s_{k'}$ & $\mathcal{L}(\cdot), \mu$ & Lagrangian, Lagrangian multiplier, of $\mathbf{P5}$ \\
					$\mathcal{N}'$,$\mathcal{N}''$ & set of PBs with transmit power smaller than, greater than or equal to, $p_\text{max}$ &
					$\bm{\zeta}_n^{\text{ap}},\bm{\zeta}_k^{\text{ue}}$ & 3D coordinate position of the $\text{PB}_n$'s antenna array $[x^\text{ap},y^\text{ap},z^\text{ap}]^T$, and $\text{UE}_k$ $[x^\text{ue},y^\text{eu},z^\text{ue}]^T$
					\\ $l_0$ & separation (m) between consecutive PBs & $d(\cdot,\cdot)$& distance function \\ 
					$\beta^\text{los}_{kn}$ & large-scale LOS fading coefficient of channel $\text{PB}_n\!\!\leftrightarrow\!\text{UE}_k$ 		
					& $\beta^\text{nlos}_{kn}$ & large-scale NLOS  fading coefficient of channel $\text{PB}_n\!\!\leftrightarrow\!\text{UE}_k$ \\
					$\mathbf{h}^\text{los}_{kn},\mathbf{h}^\text{nlos}_{kn}$ & LOS, NLOS, component of channel $\mathbf{h}_{kn}$ & $\bm{\Phi}_{kn}$ & LOS phase shift of channel $\text{PB}_n\!\leftrightarrow\!\text{UE}_k$\\
					$\theta_{kn}$ & azimuth angle relative to the boresight of the $\text{PB}_n$'s antenna array &  $\upsilon_k,\omega_k$ & angular position measured in the $x-y$ plane, $z$ axis, of a given point with respect to $\text{UE}_k$ \\
					$r_k$ & distance between a given point and $\text{UE}_k$&  
					$P_{D,k}(r_k)$ & RF power density at a distance $r_k$ from $\text{UE}_k$ \\ $\Theta_1$ & RF power density constraint &
					$P(x,y,z)$ & RF LOS receive power at point $[x,y,z]$ \\ $\varphi_{kn}$ & initial/reference phase shift of LOS channel $\text{PB}_n\leftarrow\text{UE}_k$ &
					$\Omega_0,V_0$ & sum RF power in, volume of, the complete region $R_1$ 					
					\\
					$R_0$, $R_1$ & proximity/distant region of the radio stripes & $G'$ & height of the region $R_0$ \\
					$\Theta_2,\varepsilon$ & allowable RF power exposition level at any point, and tolerable violation probability
					& $\Omega_k$ & sum RF power in, volume of, the sphere defined by the $r_k-$proximity region					
					\\				
					$a,b$ & fitting parameters of logistic EH function & $\eta$ & EH conversion efficiency\\ 
					\bottomrule		
			\end{tabular}
		\end{table*}
		\item We provide numerical evidence that training is not a very critical 
		factor for the considered indoor (short-range) system as a small amount of energy is required to attain performance similar to that of a system with ideal CSI. Moreover, we show that the system incurs in a greater transmit power consumption once the per-PB power and EMF-related constraints are considered.
		\item We discuss key trade-offs between spatial division multiple access (SDMA) and time division multiple access (TDMA). We show that TDMA is more efficient when serving a small number of UEs, while SDMA may be preferable  when the number of UEs is large.  
	\end{enumerate}
	\subsection{Organization} 
	The remainder of this paper is organized as follows. Section~\ref{system} introduces the system model and problem formulation. Section~\ref{S3} presents global and low-complexity solutions, while Section~\ref{S5} addresses the single UE case. Section~\ref{S4} introduces the analytical framework for 
	EMF-related metrics. Finally, Section~\ref{results} presents numerical results, and Section~\ref{conclusions} concludes the article.
	\newline\textit{Notation:} 
	Boldface lowercase/uppercase letters denote column vectors/matrices; $\mathbf{1}$ is a vector of ones; while $\mathbf{I}$ is the identity matrix. Superscripts $(\cdot)^*$, $(\cdot)^T$ and $(\cdot)^H$ are the complex conjugate, transpose and Hermitian operations, respectively. $||\cdot||$ and $\Tr(\cdot)$ are the Euclidean norm of a vector and trace of a matrix, respectively. The curled inequality symbol $\succeq$ denotes generalized inequality: between vectors, it represents component-wise inequality; between symmetric matrices, it represents matrix inequality. Meanwhile, $\mathbb{R}^+$ is the set of non-negative real numbers, $\mathbb{S}^d$ is the set of Hermitian matrices of dimensions $d\times d$, $\mathbb{C}$ is the set of complex numbers, respectively, and $\mathbbm{i}=\sqrt{-1}$ is the imaginary unit. Additionally, $\min\{\cdot\}$, $\{\cdot\}$ and $\lfloor\cdot\rfloor$ are the minimum, fractional part, and floor functions, respectively, while $|\cdot|$ is the absolute (or cardinality for sets) operation. $\mathbb{E}[\!\ \cdot\ \!]$ denotes statistical expectation, $\Re\{\cdot\}$ outputs the real part of the argument, and $\partial$ is partial differentiation. $\mathcal{O}(\cdot)$ is the big-O notation, and $\mathbf{x}\sim\mathcal{CN}(\bar{\mathbf{x}},\mathbf{R})$ is a circularly-symmetric complex Gaussian random vector with mean $\bar{\mathbf{x}}$ and covariance matrix $\mathbf{R}$. Table~\ref{tab1} summarizes the acronyms and main symbols used throughout this paper.
	\section{Radio Stripes System Model}\label{system}
	As in Fig.~\ref{Fig1}, we consider a mMIMO radio stripes network comprising of a set $\mathcal{N}$ of $|\mathcal{N}|=N$ PBs\footnote{Although not relevant for the scenario discussed here, we may assume that front-haul connections goes from $\text{PB}_1\rightarrow$  $\text{PB}_2\rightarrow$ $\text{PB}_3\rightarrow\cdots\rightarrow$  $\text{PB}_N\rightarrow$  central processing unit (CPU).}, each with $M$ antennas, wirelessly powering $K\le MN$ single antenna UEs in the downlink. In practice, $K\ll MN$ should hold, and  might be required so that the radio stripes system can energize relatively power hungry devices such as mobile phones, console controllers, etc. Such $K$ UEs may have been scheduled in advance for the system to charge, thus may be part of a greater set of UEs with charging demands.
		
	The PBs are equally placed on a radio stripe of length $L$ (m) which is wrapped around a square perimeter of the same length, e.g., a room or office, at a height $G$ from the floor level. Then, the number of PBs that can be deployed is upper bounded by  
	\begin{align}
	N<\frac{L}{M\lambda/2}=\frac{2L}{M\lambda}\label{N1}
	\end{align}
	by considering half-wavelength spaced antenna elements, where $\lambda$ is the system operation wavelength. Moreover, we consider quasi-static block fading, and denote the channel between $\text{PB}_n$ and $\text{UE}_k$ as $\mathbf{h}_{kn}$ and its first and second order statistics by $\mathbf{\bar{h}}_{kn}=\mathbb{E}[\mathbf{h}_{kn}]$ and $\mathbf{R}_{kn}=\mathbb{E}[\mathbf{h}_{kn}\mathbf{h}_{kn}^H]$. Here, $\mathbf{R}_{kn}\!\in\!\mathbb{C}^{M\times M}$ is the positive semi-definite channel covariance matrix. Further, the coherence block consists of $\tau$ channel uses. Prior to the downlink WET, there is an uplink channel estimation phase consisting of $\tau_p$ channel uses for pilots transmissions from the UEs. Therefore, the downlink WET occurs over the remaining $\tau\!-\!\tau_p$ channel uses. Both phases are illustrated in Fig.~\ref{Fig1} and described in the following. 
	\subsection{Uplink Channel Estimation}\label{CSI}
	We assume there are $\tau_p$ mutually orthogonal $\tau_p-$length pilot vector signals $\bm{\psi}_1, \bm{\psi}_2, \cdots , \bm{\psi}_{\tau_p}$, with $||\bm{\psi}_i||^2 = \tau_p,\ i=1,\cdots,\tau_p$, which are used by the $K$ UEs for channel estimation. We assume that $K\le \tau_p$ and the pilot sequence assigned to $\text{UE}_k$ is $\bm{\psi}_k$, thus, each UE owns a unique pilot sequence. Then, the pilot signal received at $\text{PB}_n$ in the CSI acquisition phase is 
	\begin{align}
	\mathbf{Y}_n = \sum_{k=1}^{K}\sqrt{p_k}\mathbf{h}_{kn}\bm{\psi}_{k}^T+\mathbf{W}_n,
	\end{align}
	where $p_k\ge 0$ is the fixed transmit power of $\text{UE}_k$, and $\mathbf{W}_n\in\mathbb{C}^{M\times\tau_p}$ is the noise at the receiver modeled with independent entries distributed as $\mathcal{CN}(0,\sigma^2)$. 
	
	Herein, we assume the least square (LS) channel estimate 
	\begin{align}
	\hat{\mathbf{h}}_{kn} = \frac{1}{\sqrt{p_k}\tau_p}\mathbf{Y}_n\bm{\psi}_k^*,
	\end{align}
	where the channel error estimate $\tilde{\mathbf{h}}_{kn}$ is distributed as
	\begin{align}
	\tilde{\mathbf{h}}_{kn} &= \frac{1}{\sqrt{p_k}\tau_p}\mathbf{W}_n\bm{\psi}_k^*\sim \mathcal{CN} \Big(\mathbf{0},\frac{\sigma^2}{p_k\tau_p}\mathbf{I}\Big).  
	\end{align}
	Note that more complex methods, such as the minimum mean square error estimator (MMSE), may not be needed for this kind of short-distance setup. This is because pilots are orthogonal, and $\tfrac{\sigma^2}{p_k\tau_p}\!\ll\! \mathbb{E}\big[||\mathbf{h}_{kn}||^2\big]$, thus, channel estimates are already very precise when using LS. Moreover, the MMSE method would require prior knowledge of $\bar{\mathbf{h}}_{kn}$ and $\mathbf{R}_{kn}$, which are subject to estimation errors that propagate to the actual estimate of $\mathbf{h}_{kn}$ \cite{Ozdogan.2019}.
	\subsection{Downlink WET}\label{dwwet}
	At each channel use of the downlink WET phase, all PBs transmit the same set of $K'\le K$ energy symbols $s_{k'}\in\mathbb{C}$, thus paving the way to coherent signal combinations at the UEs. The RF signal received by $\text{UE}_k$ is given by
	\begin{align}
	y_k = \sum_{n=1}^{N}\sum_{k'=1}^{K'} \mathbf{v}_{k'n}^T\mathbf{h}_{kn}s_{k'},\label{yk}
	\end{align}
	where $\mathbf{v}_{k'n}\in\mathbb{C}^{M}$ is the precoding vector associated to $s_{k'}$. Note that, in general, the number of powering signals, thus, the number of precoders, does not necessarily need to match the number of UEs. However, as the energy requirements per UE become stringent, a dedicated \emph{beam}\footnote{Here, we are not referring to concentrated beams pointing to the UEs and produced by traditional antenna arrays, but to a more disperse \textit{beam} coming from a distributed antenna array deployment. The coherent energy combination is concentrated at the UEs' position, similar to what happens in rich scattering environments (even when using traditional antenna arrays).} 
	per UE may be required. Moreover, noise impact is ignored since it is practically null for EH purposes \cite{Clerckx.2021,Clerckx.2016,LopezMontejo.2020,Lopez.2021,Rosabal.2020,Yang.2015,Khan.2018,Lee.2018,Pan.2020,Zhang.2018,XuZeng.2018,Clerckx.2018}, and it is assumed that symbols $s_{k'}$ are independent random variables, i.e., $\mathbb{E}[s_{k}^*s_{k'}]=0$, $\forall k\ne k'$, normalized so as to satisfy a unit average power constraint, i.e., $\mathbb{E}[|s_{k'}|^2]=1$. Then, the incident average RF power available at each $\text{UE}_k$ per channel use of the WET phase is given by
	\begin{align}
	P_k &= \mathbb{E}_s[|y_k|^2] = \mathbb{E}_s[y_k^*y_k]\nonumber\\
	& \stackrel{(a)}{=}   \mathbb{E}_s\Bigg[\bigg(\sum_{k_1'=1}^{K'}\!\sum_{n_1=1}^{N}\!\!\mathbf{v}_{k_1'n_1}^T\!\!\mathbf{h}_{kn_1}s_{k_1'}\!\bigg)^{\!\!*}\!\sum_{k_2'=1}^{K'}\!\sum_{n_2=1}^{N}\!\!\mathbf{v}_{k_2'n_2}^T\!\!\mathbf{h}_{kn_2}s_{k_2'}\Bigg]  \nonumber\\
	&\stackrel{(b)}{=}   \sum_{k_1'=1}^{K'}\sum_{k_2'=1}^{K'}\sum_{n_1=1}^{N}\sum_{n_2=1}^{N}\mathbf{v}_{k_1'n_1}^H\mathbf{h}_{kn_1}^*\mathbf{v}_{k_2'n_2}^T\mathbf{h}_{kn_2}\mathbb{E}_s[s_{k_1'}^*s_{k_2'}]\nonumber\\
	&\stackrel{(c)}{=} \sum_{k'=1}^{K'}\bigg|\sum_{n=1}^{N}\mathbf{v}_{k'n}^T\mathbf{h}_{kn}\bigg|^2 , \label{Pk} 
	\end{align}
	where $(a)$ comes from using \eqref{yk}, $(b)$ follows after re-arranging terms, and $(c)$ is immediately attained from leveraging the assumption of independent and power-normalized signals. Meanwhile, the energy harvested by $\text{UE}_k$ in a block (comprising both phases) converges to
	\begin{align}
	E_k = \Big(1-\frac{\tau_p}{\tau}\Big)g_k(P_k) \label{Ek}
	\end{align}
	for sufficiently large $\tau-\tau_p$, e.g., $\tau-\tau_p\gg 10$, and considering unit time blocks without loss of generality. Note that $g_k: \mathbb{R}^+\rightarrow 0\cup[\varpi_k,\nu_k]$ is a non-decreasing function modeling the relation between the incident RF power and harvested direct current (DC) power at $\text{UE}_k$, and $\varpi_k$ and $\nu_k$ are the DC sensitivity and saturation levels, respectively. Although, in general, $g_k$ also depends on the modulation and incoming waveform \cite{Clerckx.2018}, we may ignore such effects if no optimization or different modulation schemes are used as is the case here.
	\subsection{Problem Formulation}\label{PF}
	The goal is  to minimize the power consumption of the radio stripes system by properly designing/setting the precoding vectors $\{\mathbf{v}_{kn}\}$. 
	
	The system is subject to per-PB power and per-UE EH constraints at each WET phase. This is, each PB circuit is subject to a total power constraint $p_\text{max}$ (to support low-cost hardware, and limit heat dissipation), and each $\text{UE}_k$ has a target energy requirement $\xi_k$ per block. With a proper knowledge of devices EH hardware characteristics, i.e., $\{g_k\}$, the EH constraint can be transformed to 
	\begin{align}
	\hat{E}_k\ge \xi_k \rightarrow \hat{P}_k&\ge g_k^{-1}\Big(\frac{\xi_k}{1-\tau_p/\tau}\Big) \triangleq \delta_k,\label{const}
	\end{align}
	where the hat accent denotes estimation since the true channel realizations are never known with perfection, thus $\hat{E}_k$ and $\hat{P}_k$ can be evaluated using \eqref{Ek} and \eqref{Pk}, respectively, but substituting $\mathbf{h}_{kn}$ by $\hat{\mathbf{h}}_{kn}$. In practice, a constraint in the form of \eqref{const} does not prevent that the true harvested energy may fall below the threshold $\xi_k$ for poorly estimated channel realizations. However, as we shall see later in Section~\ref{results}, channel estimates are indeed very accurate in the considered setup. Moreover, small fluctuations around $\delta_k$ average out when considering consecutive coherence time intervals and will not impact the WET performance in practice.
	
	Since $g_k$ is a non-decreasing function, it is invertible as long as $\frac{\xi_k}{1-\tau_p/\tau}<\nu_k$.\footnote{Obviously $\frac{\xi_k}{1-\tau_p/\tau}>\nu_k$ is not practically viable; while in case $\frac{\xi_k}{1-\tau_p/\tau}=\nu_k$, we can set $g_k^{-1}(\nu_k)$ to be $\lim\limits_{x\rightarrow \nu_k^-}g_k^{-1}(x)$.} Therefore, the optimization problem can be formulated as
	\begin{subequations}\label{P}
		\begin{alignat}{2}
		\mathbf{P1:}\ \ &\underset{ \mathbf{v}_{k'n}\in\mathbb{C}^{M},\ {\forall k',n}}{\mathrm{minimize}}       &\ \ \ & 
		P_T = \sum_{n=1}^N \sum_{k'=1}^{K'}\!||\mathbf{v}_{k'n}||^2 \label{P1:a}\\ 
		&\text{subject to} & &  \hat{P}_k(\{\mathbf{v}_{k'n}\})\ge \delta_k,\ \ \ \ \ \forall k,  \label{P1:b}\\
		&  & &  \sum_{k'=1}^{K'}||\mathbf{v}_{k'n}||^2\leq p_\text{max},\  \forall n,  \label{P1:c}
		\end{alignat}	
	\end{subequations}
	which is non-convex due to \eqref{P1:b}. In Sections \ref{S3}-\ref{S5} we discuss different optimization approaches to solve $\mathbf{P1}$. Meanwhile, note that the radio stripes deployment topology enables a distributed 3D beamforming (avoiding the generation of strong beams since transmit antenna elements are spread over a square perimeter) pointing to the locations of the UEs, where the EMF exposure levels are expected to be the greatest. However, $\mathbf{P1}$, in its current form, does not prevent increased EMF levels at other spatial points as well. An EMF-aware optimization, under which EMF-related constraints are incorporated into $\mathbf{P1}$, is addressed later in Section~\ref{S4}.
	\section{Optimization Framework}\label{S3}
	Let us proceed by defining 
	\begin{align}
	\mathbf{v}_{k}&= [\mathbf{v}_{k1}^T, \mathbf{v}_{k2}^T,\cdots,\mathbf{v}_{kN}^T]^T,\label{vk}\\
	\hat{\mathbf{h}}_k&= [\hat{\mathbf{h}}_{k1}^T, \hat{\mathbf{h}}_{k2}^T,\cdots,\hat{\mathbf{h}}_{kN}^T]^T.\label{hk}
	\end{align}
	Then, departing from \eqref{Pk} we have that
	\begin{align}
	\hat{P}_k&=\sum_{k'=1}^{K'}\bigg|\sum_{n=1}^{N}\mathbf{v}_{k'n}^T\hat{\mathbf{h}}_{kn}\bigg|^2=\sum_{k'=1}^{K'}|\mathbf{v}_{k'}^T\hat{\mathbf{h}}_k|^2\nonumber\\
	&=\sum_{k'=1}^{K'} (\mathbf{v}_{k'}^T\hat{\mathbf{h}}_k)^H(\mathbf{v}_{k'}^T\hat{\mathbf{h}}_k)=\sum_{k'=1}^{K'}\hat{\mathbf{h}}_k^{ H}\mathbf{v}_{k'}^*\mathbf{v}_{k'}^T\hat{\mathbf{h}}_k\nonumber\\
	&=\hat{\mathbf{h}}_k^{ H}\bigg[\sum_{k'=1}^{K'}(\mathbf{v}_{k'}\mathbf{v}_{k'}^H)^T\bigg]\hat{\mathbf{h}}_k=\Tr(\mathbf{V}^T\hat{\mathbf{H}}_k), \label{Pk3}
	\end{align}
	where $\mathbf{V}=\sum_{k'=1}^{K'}\mathbf{v}_{k'}\mathbf{v}_{k'}^H$ and $\hat{\mathbf{H}}_k=\hat{\mathbf{h}}_k\hat{\mathbf{h}}_k^{ H}$.
	Now, $\mathbf{V}$ can  be  optimally found by solving the (convex) SDP in \eqref{P2}. Here, $\mathbf{B}\in \{0,1\}^{N\times NM}$ is a sensing matrix designed to collect the total transmit power of each PB, thus, it has unit entries from the $(i(M-1)+1)-$th until the $iM-$th element of each row $i=1,\cdots,N$, and zero in the remaining entries.	
	\begin{subequations}\label{P2}
		\begin{alignat}{2}
		\mathbf{P2:}\ \ \ &\underset{\mathbf{V}\in \mathbb{S}^{MN}}{\mathrm{minimize}}       &\ \ \ & 
		P_T = \Tr(\mathbf{V}) \label{P2:a}\\ 
		&\text{subject to} & &  \Tr(\mathbf{V}^T\hat{\mathbf{H}}_k)\ge \delta_k,\ \ \forall k, \label{P2:b}\\
		&  & & \ \mathbf{B}\mathrm{diag}(\mathbf{V})\preceq p_\text{max}\mathbf{1}, \label{P2:c}\\
		& & &\qquad\qquad \mathbf{V}\succeq \mathbf{0}. \label{P2:d}
		\end{alignat}	
	\end{subequations}
	\noindent 
	Notice that \eqref{P2:a}, \eqref{P2:b} and \eqref{P2:c} are equivalent to \eqref{P1:a}, \eqref{P1:b} and \eqref{P1:c}, respectively. After solving $\mathbf{P2}$ using standard convex optimization frameworks, e.g., CVX \cite{cvx}, the composite precoding vectors $\{\mathbf{v}_{k'}\}$, with $K'$ equal to the rank
	of $\mathbf{V}$, can be obtained as the eigenvectors of $\mathbf{V}$. 
	
	Interestingly, the precoding optimization for information multicasting, where the same information-bearing signal is simultaneously transmitted to all users, usually relies on a semi-definite relaxation with a structure similar to that of \eqref{P2} \cite{Sidiropoulos.2006,Tran.2014,Hsu.2017,Tervo.2018}. Therefore, after solving the SDP, it requires a last step to force rank-1 multicast beamforming, e.g., via Gaussian randomization, thus leading to sub-optimal solutions. Meanwhile, the optimum energy precoding here is not rank constrained since energy can be collected from any arbitrary superposition of energy carrying symbols, thus, directly obtained from solving $\mathbf{P2}$.
	
	Note that by removing the per-PB power constraint \eqref{P2:c} from $\mathbf{P2}$, the resulting relaxed problem is always feasible. This is because for any precoding phase design, the corresponding transmit power could boundlessly increase until constraints in \eqref{P2:b} are met. Therefore, the problem feasibility can be verified by i) finding the minimum $p_\text{max}$ that allows satisfying \eqref{P2:b}$-$\eqref{P2:d},
	and ii) verifying that it is indeed not greater than the physical per-PB power constraint. The feasibility may be controlled by the charging scheduling decision, which establishes the set of UEs to be served, thus regulating  the influence of  constraint \eqref{P2:b}.

SDP problems are solved by interior-point methods in polynomial time \cite{Hsu.2017}. 
Unfortunately, solving $\mathbf{P2}$ becomes computationally costly in the massive antenna regime.
To overcome this, low-complexity optimization approaches are presented next.
	\begin{algorithm}[t!]
		\caption{SCA-based Optimum Precoder}
		\begin{algorithmic}[1] \label{alg1-P4}
			\STATE \textbf{Input:} $\{\hat{\mathbf{h}}_{kn}\},\ \delta_k,\ p_\text{max} \ \ \forall k\in\mathcal{K}, \ \forall n\in\mathcal{N}$ \label{lin1-P4}
			\STATE Initialize $\{\overline{\mathbf{v}}_{k'n}^{}\}, \ \forall k'\in\mathcal{K}, \ \forall n\in\mathcal{N}$  \label{lin2-P4}			
			\REPEAT \label{lin3-P4}		
			\STATE Solve problem~$\mathbf{P3}$ with $\{\overline{\mathbf{v}}_{k'n}^{}\}$ and denote the local solution as $\{\mathbf{v}_{k'n}^{\star}\}$ \label{lin4-P4}
			\STATE Update $\{\overline{\mathbf{v}}_{k'n}^{}\} \leftarrow \{{\mathbf{v}_{k'n}^{\star}}\}, \ \forall k\in\mathcal{K}, \ \forall n\in\mathcal{N}$ \label{lin5-P4}		
			\UNTIL{\text{convergence or fixed number of iterations}} \label{lin8-P4}
			\STATE \textbf{Output:}  $\{{\mathbf{v}_{k'n}^{\mathrm{opt}}}\},\ \ \forall k\in\mathcal{K}, \ \forall n\in\mathcal{N}$ \label{lin9-P4}
		\end{algorithmic}
	\end{algorithm}
	\subsection{SCA-based Precoder Design}\label{subsec:SCA-beamformer}
    Herein, we adopt the SCA technique~\cite{Tran.2014,dkumar2021TWC}, wherein the non-convex problem is recast as a sequence of convex subproblems, and then iteratively solved until convergence. First, let us rewrite $\hat{P}_k$ as 
	\begin{align}
	\hat{P}_k = \sum_{k'=1}^{K}\mathbf{v}_{k'}^H\hat{\mathbf{H}}_k^T\mathbf{v}_{k'},
	\end{align}
	which comes from a procedure similar to that leading to \eqref{Pk3}. Now, observe that constraint~\eqref{P1:b} can be written as
	$\delta_k - \hat{P}_k \leq 0, \ \forall k$, where $\hat{P}_k$ is quadratic, thus a convex function~\cite[Ch. 3]{boyd2004convex}. Therefore, the rewritten constraint constitutes a difference of convex functions, thus its best convex approximation can be obtained by replacing $\hat{P}_k$ with its first-order approximation~\cite{dkumar2021TWC}. The first-order Taylor approximation of $\hat{P}_k$ around a fixed operating point~$\{\overline{\mathbf{v}}_{k'}^{}\}$ can be expressed~as 
	
	\begin{align}
	\label{eq:SCA_v}
	\widetilde{\hat{P}}_k &\triangleq \sum_{k'=1}^{K}\overline{\mathbf{v}}_{k'}^H\hat{\mathbf{H}}_k^T\overline{\mathbf{v}}_{k'}+2\sum_{k'=1}^{K}\Re\big\{\overline{\mathbf{v}}_{k'}^H\hat{\mathbf{H}}_k^T(\mathbf{v}_{k'}-\overline{\mathbf{v}}_{k'})\}  \nonumber\\
	& =  2\sum_{k'=1}^{K}\Re\big\{\overline{\mathbf{v}}_{k'}^H\hat{\mathbf{H}}_k^T\mathbf{v}_{k'}\big\} - \Tr(\overline{\mathbf{V}}^T\hat{\mathbf{H}}_k), 
	\end{align}
	 where $\overline{\mathbf{V}}=\sum_{k=1}^K\overline{\mathbf{v}}_k\overline{\mathbf{v}}_k^H$. Hence, $\mathbf{P1}$ can be approximated in the vicinity of a fixed operating point  $\{\overline{\mathbf{v}}_{k'n}^{}\}$ as 
	\begin{subequations}\label{P4}
		\begin{alignat}{2}
		\mathbf{P3:}\ \ &\underset{ \mathbf{v}_{k'n}\in\mathbb{C}^M,\ {\forall k',n}}{\mathrm{minimize}}       &\ \ \ & 
		P_T = \sum_{n=1}^N \sum_{k'=1}^{K'}\!||\mathbf{v}_{k'n}||^2 \label{P4:a}\\ 
		&\text{subject to} & &  \widetilde{\hat{P}}_k\big(\{\mathbf{v}_{k'}\},\{\overline{\mathbf{v}}_{k'}^{}\}\big) \ge \delta_k,\ \forall k,  \label{P4:b}\\
		&  & &  \sum\nolimits_{k'=1}^{K'}||\mathbf{v}_{k'n}||^2\leq p_\text{max},\ \forall n.  \label{P4:c}
		\end{alignat}	
	\end{subequations}
	Thus, by iteratively solving problem~$\mathbf{P3}$ while updating~$\{\overline{\mathbf{v}}_{k'n}^{}\}$ with the solution at each iteration, we can find the best local optimal solution~\cite{dkumar2021TWC}.  The proposed low-complexity iterative method for problem~$\mathbf{P1}$ is summarized in Algorithm~\ref{alg1-P4}.
	\subsection{MRT based Precoder Design} \label{subsec:MRT-beamformer}	
Herein, we exploit further complexity reduction by adopting an approach similar to the proposed in~\cite{Yang.2015,Khan.2018,Lee.2018,Lopez.2021}. The key lies in setting $K'=K$ and adopting a precoder design akin to MRT in MIMO communications as follows
	\begin{align}
	\mathbf{v}_{kn}= \frac{{\hat{\mathbf{h}}_{kn}^*}}{||\hat{\mathbf{h}}_{kn}||}\sqrt{p_{kn}},\qquad \forall k, n, \label{wk} 
	\end{align}
	where $p_{kn}$ represents the power budget of $\text{PB}_n$ for $s_k$.\footnote{Conjugate precoders differing only from \eqref{wk} in the normalization factor, such as the enhanced normalized conjugate beamforming (ECB) proposed and analyzed in \cite{Interdonato.2021,Sutton.2021} for multi-user cell-free mMIMO communications, may be used here as well. ECB promotes robustness against channel estimation errors because its inherent hardening capability. However, in our considered short-range indoor WET setup where channel estimation errors are practically negligible and interference is not a foe but a friend, ECB does not outperform \eqref{wk}. We opt for \eqref{wk}, which is more robust against rounding numerical errors when running the SCA-based optimization described in Algorithm~\ref{alg1-P5}.} Then, with perfect CSI knowledge, the $k-$th signal transmitted from all radio stripes antennas  arrives at $\text{UE}_k$ with constructive superposition. Meanwhile, note that the impact of the signals meant to other devices  on the energy harvested at $\text{UE}_k$ is not considered for the phases' design, thus leading to an unavoidable performance degradation with respect to the SDP-based and/or SCA-based solution previously discussed. With above precoder, the estimated RF incident power available at each $\text{UE}_k$ per channel use of a WET phase given in \eqref{Pk} transforms to
	\begin{align}
	\hat{P}_k &= \sum_{k'=1}^{K}\Bigg|\sum_{n=1}^{N} \sqrt{p_{k'n}} \frac{\hat{\mathbf{h}}_{k'n}^{H}}{||\hat{\mathbf{h}}_{k'n}||}\hat{\mathbf{h}}_{kn} \Bigg|^2\nonumber\\
	&=\sum_{k'=1}^{K}\Big|\sum_{n=1}^{N} {\varrho_{k'n}} \mathbf{q}_{k'n}^{H} \hat{\mathbf{h}}_{kn} \Big|^2, \label{Pk2} 
	\end{align}
	where $\varrho_{k'n} \triangleq \sqrt{p_{k'n}} $ becomes now the optimization variable, and $\mathbf{q}_{k'n} \triangleq \hat{\mathbf{h}}_{k'n}\big/||\hat{\mathbf{h}}_{k'n}||$ denotes the normalized precoder. 
	
	We can observe that by using expression~\eqref{Pk2} in problem~$\mathbf{P1}$, the constraint~\eqref{P1:b} is a difference of convex functions, and hence~$\mathbf{P1}$ is still non-convex problem. We again resort to SCA framework~\cite{dkumar2021TWC} to find a solution. The first-order Taylor approximation of~$\hat{P}_k$ around a fixed operating point $\{\overline{\varrho}_{k'n}^{}\} $ can be expressed as 	
	\begin{align}
	\label{eq:SCA_MRT}
	\widetilde{\hat{P}}  &\triangleq \!  \sum_{k'=1}^{K'}\!\Big|\!\sum_{n=1}^{N} \overline{\varrho}_{k'n} \mathbf{q}_{k'n}^{H} \hat{\mathbf{h}}_{kn}  \Big|^2 \!\!+\! 2 \sum_{k'=1}^{K}\! \Re \Bigg\{\!\sum_{n=1}^{N}  \overline{\varrho}_{k'n}  \hat{\mathbf{h}}_{kn}^{H} \mathbf{q}_{k'n}\times\nonumber\\
		&\qquad\qquad\qquad \times \bigg( \sum_{n=1}^{N} \Big({\varrho}_{k'n} - \overline{\varrho}_{k'n} \Big) \mathbf{q}_{k'n}^{H} \hat{\mathbf{h}}_{kn} \bigg) \Bigg\}.
	\end{align}
	Hence, by using \eqref{eq:SCA_MRT}, the MRT precoder can be found by  solving
	\begin{subequations}\label{P5}
		\begin{alignat}{2}
		\mathbf{P4:}\ \ \ &\underset{\varrho_{kn}\in\mathbb{R}^+,\ {\forall k,n}}{\mathrm{minimize}}       &\ \ \ & 
		P_T = \sum\nolimits_{n=1}^N \sum\nolimits_{k=1}^{K} \varrho_{kn}^{2} \label{P5:a}\\ 
		&\text{subject to} & &  \widetilde{\hat{P}}\big(\{\varrho_{k'n}\},\{\overline{\varrho}_{k'n}^{}\}\big) \geq \delta_k ,\ \forall k, \label{P5:b}\\
		&  & &  \sum\nolimits_{k=1}^{K} \varrho_{kn}^{2}\leq p_\text{max},\ \ \forall n, \label{P5:c}
		\end{alignat}	
	\end{subequations}
	where the fixed operating point $\{\overline{\varrho}_{k'n}^{}\}$ is iteratively updated as summarized in Algorithm~\ref{alg1-P5}.

	Finally, note that $\mathbf{P3}$ and $\mathbf{P4}$ can be efficiently solved as a sequence of second-order cone programs (SOCPs)~\cite{boyd2004convex}. 	
	Interior points methods are adopted to efficiently solve SOCP formulations in polynomial time with lower complexity than SDP formulations \cite{Tran.2014}. Nevertheless, observe that the number of variables, thus the complexity, is inferior in case of $\mathbf{P4}$.  
	\begin{spacing}{0.8}
		\begin{algorithm}[t!]
			\caption{MRT-based Precoder}
			\begin{algorithmic}[1] \label{alg1-P5}
				\STATE \textbf{Input:} $\{\hat{\mathbf{h}}_{kn}\},\ \delta_k,\ p_\text{max} \ \ \forall k\in\mathcal{K}, \ \forall n\in\mathcal{N}$ \label{lin1-P5}
				\STATE Set $i=1$ and initialize $\overline{\varrho}_{k'n}^{} \ \ \forall k'\in\mathcal{K}, \ \forall n\in\mathcal{N}$  \label{lin2-P5}			
				\REPEAT \label{lin3-P5}		
				\STATE Solve problem~$\mathbf{P4}$ with $\{\overline{\varrho}_{k'n}^{}\}$ and denote the local solution as $\{\varrho_{k'n}^{\star}\}$ \label{lin4-P5}
				\STATE Update $\{\overline{\varrho}_{k'n}^{}\} \leftarrow \{{\varrho_{k'n}^{\star}}\}, \ \forall k\in\mathcal{K}, \ \forall n\in\mathcal{N}$ \label{lin5-P5}
				\STATE Set $i = i+1$ \label{lin6-P5}	
				\UNTIL{\text{convergence or fixed number of iterations}} \label{lin8-P5}
				\STATE \textbf{Output:}  $\{{\varrho_{k'n}^{\mathrm{opt}}}\},\ \ \forall k\in\mathcal{K}, \ \forall n\in\mathcal{N}$ \label{lin9-P5}
			\end{algorithmic}
		\end{algorithm}
	\end{spacing}
	\section{Single UE Case}\label{S5}
	Herein, we study a scenario of particular interest: the single UE case. Note that the system may serve the UEs in a TDMA fashion, in which case the derivations and performance analysis here hold accurate. Alternatively, the performance results for this scenario correspond to upper bounds of what it is expected if more UEs are simultaneously served in the same time interval.
	\subsection{Optimum Precoder}
	In case of a single UE, the MRT-like precoder discussed in Section~\ref{subsec:MRT-beamformer} is the optimum. Then, let us relax the per-PB power constraint. The optimum concatenated precoder is 
	\begin{align}
	\mathbf{v}_1 = \frac{\hat{\mathbf{h}}_1^*}{||\hat{\mathbf{h}}_1||}\sqrt{p},\label{v1}
	\end{align}
	where $\mathbf{v}_1$ and $\hat{\mathbf{h}}_1$ obey \eqref{vk} and \eqref{hk}, respectively. Then, we have that $\hat{P}_1 = |\mathbf{v}_1^T\hat{\mathbf{h}}_1|^2=p||\hat{\mathbf{h}}_1||^2$, thus $p^\text{opt}=\delta_1/||\hat{\mathbf{h}}_1||^2$.
	However, such solution may demand a transmit power above the per-PB power constraint $p_\text{max}$ at some PB since $p_{1n}$ would need to be
	\begin{align}
	p_{1n} = ||\mathbf{v}_{1n}||^2= \frac{p||\hat{\mathbf{h}}_{1n}||^2}{||\hat{\mathbf{h}}_{1}||^2}\label{p1n}
	\end{align}
	and has not been bounded.
	
	Let us assume that the set $\mathcal{N}'\subseteq \mathcal{N}$ of PBs operating with transmit power below $p_\text{max}$ is known beforehand. Then, we have that
	\begin{align}
	\hat{P}_1 \!=\! \Big(&\sqrt{p_\text{max}}\!\!\sum_{n\in\mathcal{N}'\backslash\mathcal{N}}\!\!\!\!||\hat{\mathbf{h}}_{1n}||\!+\!\!\sum_{n\in\mathcal{N}'}\!\!\varrho_{n}||\hat{\mathbf{h}}_{1n}||\Big)^2\!\ge\! \delta_1 \nonumber\\
	&\rightarrow \! \sum_{n\in\mathcal{N}'}\!\!\varrho_n||\hat{\mathbf{h}}_{1n}||\! \ge\! \underbrace{\sqrt{\delta_1} \!-\! \sqrt{p_\text{max}}\!\!\sum_{n\in\mathcal{N}\backslash\mathcal{N}'}\!\!\!\!||\hat{\mathbf{h}}_{1n}||}_{\sqrt{\delta_1'}},\label{delta1}
	\end{align}
	where $\varrho_{n}\triangleq \sqrt{p_{1n}}$. Using this result, one can transform $\mathbf{P1}$ to obtain
	\begin{spacing}{0.85}
		\begin{subequations}\label{P6}
			\begin{alignat}{2}
			\mathbf{P5:}\ \ \ &\underset{\varrho_{n}\in\mathbb{R}^+,\ {\forall n}}{\mathrm{minimize}}       &\ \ \ & 
			P_T = \sum_{n\in \mathcal{N}'}  \varrho_{n}^{2} \label{P6:a}\\ 
			&\text{subject to} & &  \hat{P}_1'= \sum_{n\in\mathcal{N}'}\varrho_{n}||\hat{\mathbf{h}}_{1n}|| \geq \sqrt{\delta_1'}.
			\end{alignat}	
		\end{subequations}
	\end{spacing}
	Now, let us write the Lagrangian of $\mathbf{P5}$ as
	\begin{align}
	\mathcal{L}(\varrho_n,\mu) = \sum_{n\in\mathcal{N}'}\varrho_{n}^2+\Big(\sqrt{\delta_1'}-\sum_{n\in\mathcal{N}'}\varrho_{n}||\hat{\mathbf{h}}_{1n}||\Big)\mu,\label{L}
	\end{align}
	where $\mu\ge 0$ is the Lagrangian multiplier. Observe that \eqref{L} is a convex quadratic function of every $\varrho_n$, thus minimized at
	\begin{align}
	\frac{\partial \mathcal{L}(\varrho_n,\mu)}{\partial \varrho_{n}} = 2\varrho_{n}-\mu||\hat{\mathbf{h}}_{1n}|| &= 0 \nonumber\\
    \rightarrow \  \varrho_{n}^\text{opt} &= \frac{\mu}{2}||\hat{\mathbf{h}}_{1n}||.\label{varrho}
	\end{align}
	Then, the Lagrange dual of $\mathbf{P5}$ is given by
	\begin{subequations}\label{P7}
		\begin{alignat}{2}
		\mathbf{P6\!:}\ \underset{\mu\ge 0}{\mathrm{maximize}}\ 
		\mathcal{L}(\varrho_n^\text{opt},\mu)\!=\! \mu\sqrt{\delta_1'} \!-\!\frac{\mu^2}{4}\sum_{n\in \mathcal{N}'}||\hat{\mathbf{h}}_{1n}||^2, \label{P7:a}
		\end{alignat}	
	\end{subequations}
	for which
	\begin{align}
	\mu^\text{opt} = \frac{2\sqrt{\delta_1'}}{\sum_{n\in \mathcal{N}'}||\hat{\mathbf{h}}_{1n}||^2}. \label{mu}
	\end{align}
	Therefore, substituting \eqref{mu} into \eqref{varrho} yields the optimum solution of $\mathbf{P5}$, which is given by 
	\begin{align}
	\varrho_{n}^\text{opt} = \frac{||\hat{\mathbf{h}}_{1n}||\sqrt{\delta_1'}}{\sum_{n\in \mathcal{N}'}||\hat{\mathbf{h}}_{1n}||^2}. \label{varrhoOpt}
	\end{align}
	This means that if the set $\mathcal{N}\backslash\mathcal{N}'$ of PBs are known to operate with transmit power $p_\text{max}$, the remaining PBs optimum transmit power is given by $p_{1n}^\text{opt}=(\varrho_{n}^\text{opt})^2$, where $\varrho^\text{opt}$ is given in \eqref{varrhoOpt}.
	
	Finally, the optimum precoder can be constructed as shown in Algorithm~\ref{alg1}. In a nutshell, the transmit power of each PB is computed by exploiting \eqref{varrhoOpt} (line~\ref{lin7}). In case a power allocation exceeds the maximum allowed power $p_\text{max}$, it is immediately reduced to match  $p_\text{max}$ (lines \ref{lin4}-\ref{lin6}). The process is repeated until all PBs are allocated a transmit power not greater than $p_\text{max}$. Note that this is an extremely simple optimization algorithm, where at most $N$ iterative computations of \eqref{varrhoOpt} are required, but still provides the optimum WET precoder.
	\begin{algorithm}[t!]
		\caption{Single-UE Optimum Precoder}
		\begin{algorithmic}[1] \label{alg1}
			\STATE \textbf{Input:} $\{\hat{\mathbf{h}}_{1n}\}_{n\in\mathcal{N}},\ \delta_1,\ p_\text{max}$ \label{lin1}
			\STATE Set $\mathcal{N}'=\mathcal{N}$, $p_{1n}=0, \forall n\in \mathcal{N}$ \label{lin2}			
			\REPEAT \label{lin3}		
			\STATE Set $\mathcal{N}''=\{n\in \mathcal{N}'| p_{1n}\ge p_\text{max}\}$ \label{lin4}
			\STATE Update $p_{1n}\leftarrow p_\text{max},\ \forall n\in\mathcal{N}''$ \label{lin5}
			\STATE Update $\mathcal{N}'\leftarrow \mathcal{N}'\backslash\mathcal{N}''$ \label{lin6}
			\STATE Update $\delta_1'$ according to \eqref{delta1} \label{lin6p5}
			\STATE Set $p_{1n}=(\varrho_{n}^\text{opt})^2, \forall n\in\mathcal{N}'$, according to \eqref{varrhoOpt} \label{lin7}			
			\UNTIL{$p_{1n}\le p_\text{max},\ \forall n\in\mathcal{N}$} \label{lin8}
			\STATE \textbf{Output:}  $p_{1n}^\text{opt}\leftarrow p_{1n},\ \forall n\in\mathcal{N}$ \label{lin9}
		\end{algorithmic}
	\end{algorithm}
	
	The feasibility of a single UE optimization problem is rather simple to verify. It is enough setting $p_{1n}=p_\text{max},\forall n\in\mathcal{N}$, and determine whether $\hat{P}_1\ge \delta_1$ holds or not. In this case, $\hat{P}_1$ can be evaluated departing from \eqref{Pk2} as $\hat{P}_1=\big(\sum_{n=1}^{N}\sqrt{p}_\text{max}||\hat{\mathbf{h}}_{1n}||\big)^2=p_\text{max}\big(\sum_{n=1}^{N}||\hat{\mathbf{h}}_{1n}||\big)^2$. Then, if $\sum_{n=1}^{N}||\hat{\mathbf{h}}_{1n}||\ge \sqrt{\delta_1/p_\text{max}}$, we can conclude the problem is feasible and the solution can be obtained from Algorithm~\ref{alg1}. Finally, note that in the scenario with $K>1$ UEs it is sufficient (but not necessary) that $\sum_{n=1}^{N}||\hat{\mathbf{h}}_{kn}||< \sqrt{\delta_k/p_\text{max}}$ for some $k$ to declare that $\mathbf{P4}$ is not feasible. 
	\subsection{Minimum Average Radio Stripes Transmit Power}
	One interesting question is how much average power is required to serve the UE. The minimum average transmit power required by the radio stripes system can be computed by assuming no per-PB power constraint as follows
	\begin{align}
	\mathbf{E}[P_T]&=\mathbf{E}[\delta_1/||\hat{\mathbf{h}}_1||^2]\nonumber\\
	&\stackrel{(a)}{\ge} \delta_1/\mathbf{E}[||\hat{\mathbf{h}}_1||^2]  = \delta_1/\sum_{n=1}^{N}\Tr(\mathbf{R}_{1n}),
	\end{align}
	where $(a)$ comes from applying Jensen's inequality.
	\section{EMF-Aware Optimization}\label{S4}
\begin{figure}
		\centering
		\includegraphics[width=0.46\textwidth]{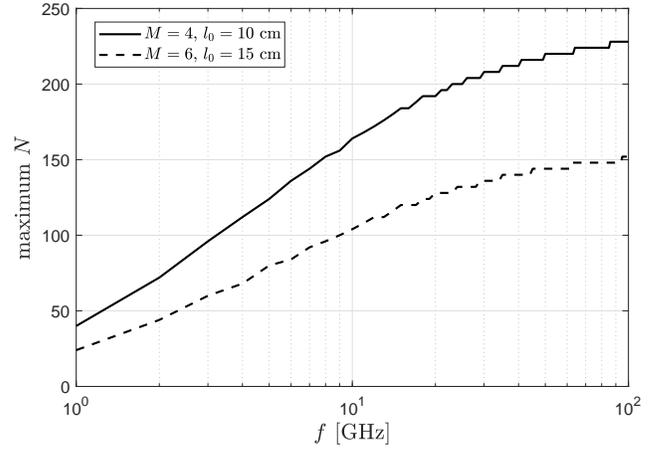}	
		\caption{Maximum number of PBs that can fit in the radio stripes system as a function of $f$. We set $L=24$~m.}
		\label{Fig2}
	\end{figure}

	In this section, we present an analytical framework to determine, analyze and limit EMF exposure levels. We assume each PB of the radio stripes (Fig.~\ref{Fig1}) is equipped with a relative small number of antennas such that $p_{\max}M$ is bounded to a desired level\footnote{Assuming a small number of antennas per PB also implies that each UE is in the far field with respect to all PBs.}, and we impose a separation restriction $l_0>\lambda/2$ between consecutive PBs. To ease the deployment and system design, we assume that no PB expands over any consecutive two sides of the perimeter, thus, the upper bound in \eqref{N1} further tightens to 
	\begin{align}
	N\le 4\Big\lfloor\frac{L/4}{(M-1)\lambda/2+l_0}\Big\rfloor = 4\Big\lfloor\frac{L}{2(M-1)\lambda+4l_0}\Big\rfloor. \label{N2}
	\end{align}
	As an example, Fig.~\ref{Fig2} illustrates the maximum number of PBs that could be equipped in the radio stripes system as a function of the central operation frequency $f$.
	
	The channel coefficients can be decomposed as $\mathbf{h}_{kn}=\mathbf{h}_{kn}^\text{los}+\mathbf{h}_{kn}^\text{nlos}$, where addends correspond to the LOS and non-LOS (NLOS) channel components.
	Note that the focus of our work is on indoor scenarios where LOS channels are usually predominant, i.e., $||\mathbf{h}_{kn}^\text{los}||^2\gg ||\mathbf{h}_{kn}^\text{nlos}||^2$. Thus, we next characterize the LOS channels and then introduce the EMF exposure metrics.
	\subsection{LOS Geometry}\label{LOSgeom}
	The geometry of the discussed scenario is illustrated in Fig.~\ref{Fig3}. The LOS (geometric) channel component between $\text{PB}_n$ and $\text{UE}_k$ is given by
	\begin{align}
	\mathbf{h}^\text{los}_{kn} = \sqrt{\beta_{kn}^\text{los}} e^{\mathbbm{i}(\varphi_{kn}\bm{1}+\bm{\Phi}_{kn})}, \label{hlos}
	\end{align}
	where $\beta_{kn}^\text{los}$ denotes its large-scale fading component.  Herein we adopt the following 3GPP indoor LOS path-loss model \cite{3GPP} 
	\begin{align}
	\beta_{kn}^\text{los}\!=\! - 17.3\log_{10}d\big(\bm{\zeta}^{\text{ap}}_n,\bm{\zeta}^{\text{ue}}_k\big) - 20\log_{10}f \!-\! 32.4\ [\text{dB}],\label{beta} 
	\end{align}
	where $\bm{\zeta}^{\text{ap}}_n=[x^{\text{ap}}_n,y^{\text{ap}}_n,z^\text{ap}]^T$ and $\bm{\zeta}^{\text{ue}}_k=[x^{\text{ue}}_k,y^{\text{ue}}_k,z^{\text{ue}}_k]^T$ are the 3D coordinate position of the $\text{PB}_n$'s antenna array and $\text{UE}_\text{k}$, 
	$f$ is given in GHz, and distance $d(\bm{\zeta}_1,\bm{\zeta}_2)=||\bm{\zeta}_1-\bm{\zeta}_2||$ is measured in meters. Meanwhile, the $i-$th entry of $\bm{\Phi}_{kn}$ in \eqref{hlos} corresponds to the mean phase shift between the $i-$th and the first array element of $\text{PB}_n$, thus, we have \cite[Ch.~5]{Hampton.2014}
	\begin{align}
	\bm{\Phi}_{kn} = -[0,\ 1,\ \cdots,\ M-1]^T\pi\sin\theta_{kn},
	\end{align}
	where $\theta_{kn}$ is the azimuth angle relative to the boresight of the $\text{PB}_n$'s antenna array, thus
	\begin{align}
	\sin \theta_{kn} = \frac{|\varsigma^{\text{ap}}_n-\varsigma^{\text{ue}}_k|}{d_{xy}(\bm{\zeta}^{\text{ap}}_n,\bm{\zeta}^{\text{ue}}_k)},
	\end{align}
	where $\varsigma=x$ or $\varsigma=y$ if the antenna array of the corresponding PB is aligned with the $x$ or $y$ axis, respectively, while $d_{xy}(\bm{\zeta}_1,\bm{\zeta}_2)=\sqrt{(x_1-x_2)^2+(y_1-y_2)^2}$ is the $x-y$ distance component between $\bm{\zeta}_1$ and $\bm{\zeta}_2$. 
	Finally, $\varphi_{kn}$ accounts for an initial phase shift, which in our case depends on the PB and UE position as
	\begin{align}
	\varphi_{kn} =2\pi\Big\{\frac{1}{\lambda}||\bm{\zeta}_1-\bm{\zeta}_2||\Big\}.
	\end{align}
	\subsection{EMF Level in the Proximity of the UEs}\label{prox}
	Herein, we analyze the EMF exposure level for the user holding the device. In the proximity of the UEs, high EMF exposure levels may come from the coherent combination of the signal, especially if channels have not been accurately estimated. 
	In this case, an appropriate performance figure could be how fast does the RF power density falls as a spatial point moves away from the device antenna. 
	\begin{figure}
		\centering
		\includegraphics[width=0.4\textwidth]{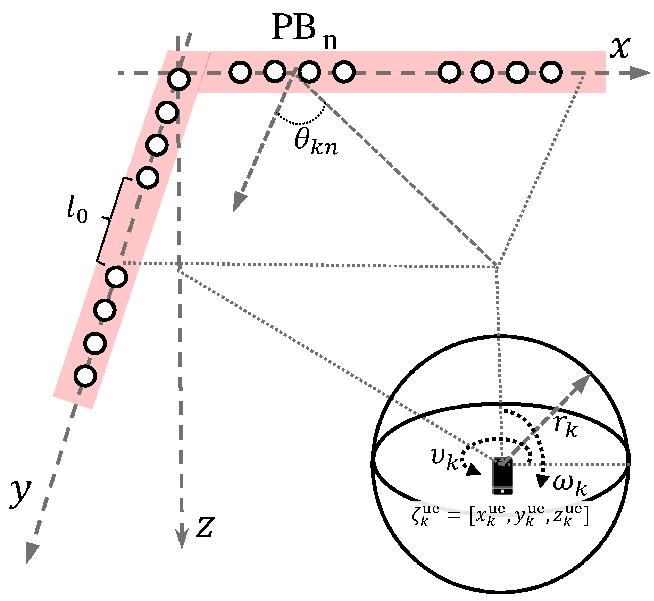}	
		\caption{Illustration of the geometry of the scenario.}
		\label{Fig3}
	\end{figure}
	
	Using the derivations in previous section, the first-order statistics of the channel at a distance $r_k$ from $\text{UE}_k$, and with angular position defined by $0\le \upsilon_{k}\le 2\pi$ (measured in the $x-y$ plane) and $0\le \omega_{k}\le \pi$ (measured in the $z$ axis), can be defined as $\mathbf{h}^\text{los}_{kn}(\upsilon_{k},\omega_{k},r_k)$ in a similar way to \eqref{hlos} but with a virtual measurement point located at 
	\begin{align}
	\bm{\zeta}_k(\upsilon_{k},\omega_{k},r_k)\! =\! [x_k^\text{ue}\!+\!r_k\sin\omega_{k}\cos&\upsilon_{k},
	y_k^\text{ue}\!+\!r_k\sin\omega_{k}\sin\upsilon_{k},\nonumber\\
	& z_k^\text{ue}+r_k\cos\omega_{k}]. 
	\end{align}
	This point lies in the $r_k-$radius sphere centered at $\zeta_k^\text{ue}$ as illustrated in Fig.~\ref{Fig3}.
	Then, the expected RF power density at a distance $r_k$ from $\text{UE}_k$, herein called $r_k-$proximity region, is given by 
	\begin{align}
	P_{D,k}(r_k)&= \frac{1}{4\pi r_k^2}\int\limits_{0}^{2\pi}\int\limits_{0}^{\pi} \Tr\big(\mathbf{V}^T\mathbf{H}^\text{los}_{k}(\upsilon_{k},\omega_{k},r_k)\big) \mathrm{d}\omega_{k}\mathrm{d}\upsilon_{k}\nonumber\\
	&= \frac{1}{4\pi r_k^2}\Tr\big(\mathbf{V}^T\bar{\mathbf{H}}_k(r_k)\big)\ \ [\text{W}/\text{m}^2], \label{PDk}
	\end{align}
	where $\bar{\mathbf{H}}_k(r_k)\!=\!\int_{0}^{2\pi}\int_{0}^{\pi}\mathbf{H}^\text{los}_{k}(\upsilon_{k},\omega_{k},r_k)\mathrm{d}\omega_{k}\mathrm{d}\upsilon_{k}$,  $\mathbf{H}^\text{los}_{k}(\upsilon_{k},\omega_{k},r_k)\!=\!\mathbf{h}^\text{los}_{k}(\upsilon_{k},\omega_{k},r_k)\mathbf{h}^\text{los}_{k}(\upsilon_{k},\omega_{k},r_k)^H$ and
	\begin{align}
	\mathbf{h}^ \text{los}_{k}(\upsilon_{k},\omega_{k},r_k)=\Big[\mathbf{h}^\text{los}_{k1}(\upsilon_{k}&,\omega_{k},r_k)^T,  \mathbf{h}^\text{los}_{k2}(\upsilon_{k},\omega_{k},r_k)^T,\nonumber\\
	&\cdots,\mathbf{h}^\text{los}_{kN}(\upsilon_{k},\omega_{k},r_k)^T\!\Big]^T.
	\end{align}
	Thus, $\bar{\mathbf{H}}_k(r_k)\succeq 0,\ \forall k$. Note that this requires using  \eqref{hlos} but with $\bm{\zeta}_k(\upsilon_{k},\omega_{k},r_k)$ instead of $\bm{\zeta}_k^\text{ue}$.
	
	Using \eqref{PDk}, the RF power density falls with a rate of
	\begin{align}
	\partial& P_{D,k}(r_k)\nonumber\\
	&= \!\frac{1}{4\pi r_k^2}\Big(\!\Tr\big(\mathbf{V}^T\partial\bar{\mathbf{H}}_k(r_k)\big)\!-\!\frac{2}{ r_k}\!\Tr\big(\mathbf{V}^T\bar{\mathbf{H}}_k(r_k)\big)\Big).\label{dP}
	\end{align}
	However, current regulations and EMF-related constraints are more compatible with $P_{D,k}(r_k)$ rather than with $\partial P_{D,k}(r_k)$. Thus, we adopt EMF constraints of the form
	\begin{align}
	P_{D,k}(r_0)\le \Theta_1,\ \ \forall k, \label{conPD}
	\end{align}
	for a certain measurement distance $r_0>0$ of interest. Then, \eqref{conPD} can be easily incorporated to the SDP and low-complexity optimization frameworks in Section~\ref{S3}. 
	\subsection{EMF Exposure Level for a Random Human}
	Herein, we set system constraints to prevent high EMF exposure to any random human (and living species), not necessarily associated to (in the proximity of) the served UEs. Note that the RF LOS receive power at a virtual measurement point $\bm{\zeta}=[x,y,z]$ is given by
	\begin{align}
	P(x,y,z)=\Tr\big(\mathbf{V}^T\mathbf{H}^\text{los}(x,y,z)\big),\label{Pxyz}
	\end{align}
	where $\mathbf{H}^\text{los}(x,y,z)=\mathbf{h}^\text{los}(x,y,z)\mathbf{h}^\text{los}(x,y,z)^H$ with
	\begin{align}
	\mathbf{h}^\text{los}(x,y,z)=\Big[\mathbf{h}^\text{los}_{k1}(x,y,z)^T,&  \mathbf{h}^\text{los}_{k2}(x,y,z)^T,\nonumber\\
	&\cdots,\mathbf{h}^\text{los}_{kN}(x,y,z)^T\Big]^T,\label{newh}
	\end{align}
	which requires using  \eqref{hlos} but with $\bm{\zeta}$ instead of $\bm{\zeta}_k^\text{ue}$. Ideally, one would like to have a constraint of the kind $P(x,y,z)\le\Theta_2,\ \forall x,y,z$ out of the $r_k-$proximity region of each $\text{UE}_k$, where $\Theta_2$ is the allowed RF power exposition level. However, this would require imposing an infinite number of constraints since the number of triplets $[x,y,z]$ is infinite, and even a discretization of the volumetric space may not be advisable as some critical points may be left out. Instead, we adopt a probabilistic constraint as
	\begin{align}
	\mathbb{P}\big[P(x,y,z)\ge\Theta_2\big]\le \varepsilon,
	\end{align}
	where $\varepsilon\ll 1$ is a tolerable violation probability\footnote{EMF-related stochastic constraints may be unavoidable in practical systems due to the randomness triggered by many known and unknown factors \cite{Wiart.2016}. 
	In the context of WET, authors in \cite{DaiZhao.2018} even propose the notion of probabilistic EMF safety that requires the probability that EMF intensity anywhere does not exceed a given threshold with a given strict confidence.}, and $x,y,z$ are  uniformly distributed random variables in the region of interest. This kind of constraint is in general difficult to address since the distribution of $P(x,y,z)$ depends on the system configuration parameters, thus, herein we transform it into an average constraint by using Markov inequality, thus, further constraining the feasibility solution set, as
	\begin{align}
	\mathbb{P}\big[P(x,y,z)\ge\Theta_2\big]\le \mathbb{E}\big[P(x,y,z)\big]/\Theta_2 &\le \varepsilon\nonumber\\
	 \rightarrow\ 
	\mathbb{E}\big[P(x,y,z)\big] &\le \varepsilon \Theta_2. \label{rest2}
	\end{align}
	
\begin{figure}
		\centering
		\includegraphics[width=0.48\textwidth]{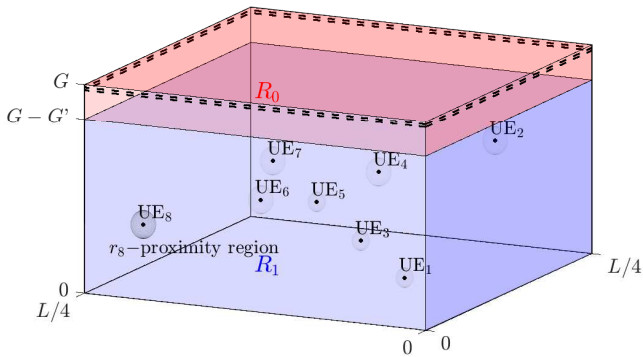}
		\caption{Main operation regions. For illustration purposes, the position of the EH devices are set as $\bm{\zeta}_k^\text{ue}=[Lk/32,Lk/32,2^{k/7-1}]$ for odd $k$, while $\bm{\zeta}_k^\text{ue}=[L(9-k)/32,Lk/36,2^{1-k/7}]$ for even $k$, with $K=8$.}
		\label{Fig4}
	\end{figure}

	To proceed further, we distinguish two main regions as illustrated in Fig.~\ref{Fig4}: 
	\begin{enumerate}
		\item[$R_0:$] proximity to the radio stripes, so that the probability of human (living) tissue goes to $0$, 
		\item[$R_1:$] remaining deployment area where humans (and living species) can freely and safely be, excluding all the $r_k-$proximity regions.
	\end{enumerate}
	We focus our discussions on $R_1$, which is critical. However we would like to place some comments regarding $R_0$, which may be a region with non-controlled or partially controlled EMF exposure levels. In case of the former, a manual switch-off mechanism may be needed if a human enters the region. Meanwhile, in case of the latter, sensing mechanisms may be deployed to detect the human presence and switch-off the system, or alternative force the system to comply with the EMF regulations also in that region. Thus, in addition to aforementioned power density and RF power exposure constraint, the system may need to comply with additional regulations when a human is sensed in $R_0$, e.g., specific absorption rate (SAR) constraints\footnote{SAR measures the absorbed power from fields between 100 kHz and 10 GHz in a unit mass of human tissue by using units of [W/kg], and dominates the RF exposure for very short distances (e.g., less than few tens of centimeters) \cite{Hochwald.2014,Zhang.2020}.}.
	
	Regarding the region of interest, $R_1$, we can proceed as follows
	\begin{align}
	\mathbb{E}\big[P(x,y,z)\big] & = \frac{\Omega_\text{0}-\sum_{k=1}^K\Omega_k}{V_\text{0}-\sum_{k=1}^KV_k},\label{EP}
	\end{align}
	where $\Omega_\text{0}$ corresponds to the sum RF power in the complete region $R_1$, thus, with volume $V_\text{0}=(G-G')L^2/16$ according to Fig.~\ref{Fig4}, and $\Omega_k$ corresponds to the sum RF power in the sphere defined by the $r_k-$proximity region, thus with volume $V_k=4\pi r_k^3/3$. Notice that
	\begin{align}
	\Omega_\text{0} &=\int\limits_{0}^{G-G'}\int\limits_{0}^{L/4}\int\limits_{0}^{L/4}P(x,y,z)\mathrm{d}x\mathrm{d}y\mathrm{d}z=\Tr\big(\mathbf{V}^T\bar{\bar{\mathbf{H}}}\big),\label{t1}\\
	\Omega_k&=\int_{0}^{r_k}\Tr\big(\mathbf{V}^T\bar{\mathbf{H}}_k(r)\big)\mathrm{d}r=\Tr\big(\mathbf{V}^T\bar{\bar{\mathbf{H}}}_k\big),\label{t2}
	\end{align}
	which come from using  $\bar{\bar{\mathbf{H}}}_k=\int_{0}^{r_k}\bar{\mathbf{H}}_k(r)\mathrm{d}r$, where  $\bar{\mathbf{H}}_k(r)$ is defined after \eqref{PDk}, and from using $P(x,y,z)$  given in \eqref{Pxyz} and setting $\bar{\bar{\mathbf{H}}}=\int_{0}^{G-G'}\int_{0}^{L/4}\int_{0}^{L/4}\mathbf{H}^\text{los}(x,y,z)\mathrm{d}x\mathrm{d}y\mathrm{d}z$. By substituting \eqref{t1} and \eqref{t2} into \eqref{EP}, \eqref{rest2} can be rewritten as
	\begin{align}
	\Tr\big(\mathbf{V}^T\mathbf{H}_\text{eff}\big) \le \varepsilon \bigg(V_0-\sum_{k=1}^KV_k\bigg) \Theta_2\triangleq \bar{\Theta}_2,\label{const2}
	\end{align}
	where $\mathbf{H}_{\text{eff}}=\bar{\bar{\mathbf{H}}}-\sum_{k=1}^K\bar{\bar{\mathbf{H}}}_k$. Then, \eqref{const2} can be immediately incorporated to the SDP and MRT-based optimization frameworks in Sections~\ref{S3} and \ref{subsec:MRT-beamformer}, respectively. 
	However, observe that $\mathbf{H}_\text{eff}$ may not be positive semi-definite, thus 
	\begin{align}
	\Tr\big(\mathbf{V}^T\mathbf{H}_\text{eff}\big)=\sum_{k=1}^{K}\mathbf{v}_k^H\mathbf{H}_\text{eff}^T\mathbf{v}_k&\le \bar{\Theta}_2
	\end{align}
	is not guaranteed to impose a convex constraint on $\{\mathbf{v}_k\}$, and cannot be incorporated directly to the SCA-based precoding designed in Section~\ref{subsec:SCA-beamformer}, and specifically given in Algorithm~\ref{alg1}. Therefore, we compute the eigendecomposition of $\mathbf{H}_\text{eff}$, which allows us to write $\mathbf{H}_\text{eff}=\mathbf{H}_\text{eff,1}+\mathbf{H}_\text{eff,2}$, where $\mathbf{H}_\text{eff,1}\succeq 0$ and $\mathbf{H}_\text{eff,2}\prec 0$ are the matrices associated with the positive and negative eigenvectors of $\mathbf{H}_\text{eff}$, respectively. Since now $\Tr\big(\mathbf{V}^T\mathbf{H}_\text{eff}\big)=\Tr\big(\mathbf{V}^T\mathbf{H}_\text{eff,1}\big)+\Tr\big(\mathbf{V}^T\mathbf{H}_\text{eff,2}\big)$ is the sum of a convex and concave function, one can easily apply the SCA technique here. Specifically, the linearization is applied to the (concave) second term, thus, similar to \eqref{eq:SCA_v}, the additional iterative constraint to be included in $\mathbf{P3}$ is given by 
	\begin{align}
	\Tr\big(\mathbf{V}^T\mathbf{H}_\text{eff,1}\big) \!+\! 2\sum_{k=1}^{K}\Re\big\{\overline{\mathbf{v}}_k^H\mathbf{H}_{\text{eff,2}}^T\mathbf{v}_k\big\} \!-\! \Tr(\overline{\mathbf{V}}^T\mathbf{H}_{\text{eff,2}})\!\le\! \bar{\Theta}_2.
	\end{align}

	Finally, note that due to the use of Markov inequality in the derivations, the violation probability of $P(x,y,z)\le \Theta_2$ is in fact smaller than $\varepsilon$, but cannot be established beforehand.
	\section{Numerical Results}\label{results}
	In this section, we illustrate the performance of discussed radio stripes system for WET with/without the introduced EMF-related constraints.
	\subsection{Configuration setup}\label{conf}
	Unless stated otherwise we set the system parameters as given next.
	\paragraph{Scenario Geometry}
	We consider a room with dimensions $6\times 6\times 3\ \mathrm{m}^3$, where the radio stripes system is assumed at the ceiling level, i.e., $z^\text{ap}=G=3$ m. Moreover, we set $G'=2.5$ m, and unless stated otherwise, $f=4$ GHz is the central operation frequency, for which $\lambda = 7.5$ cm. The minimum separation constraint between consecutive PBs is set to $l_0 = 10$ cm.
	\paragraph{Channel Model}
	We assume channels experience Rician quasi-static fading, and remain static for $\tau=1000$ channel uses. Note that under this distribution assumption, it holds that $\bar{\mathbf{h}}_{kn}=\mathbf{h}^\text{los}_{kn}$, which is given in \eqref{hlos}, and 
		$\mathbf{h}_{kn}^\text{nlos}\sim\mathcal{CN}\big(\mathbf{0},\mathbf{R}_{kn}-\bar{\mathbf{h}}_{kn}\bar{\mathbf{h}}_{kn}^H\big)$, where
	\begin{align}
	\Tr(\mathbf{R}_{kn}-\bar{\mathbf{h}}_{kn}\bar{\mathbf{h}}_{kn}^H)&=M\beta_{kn}^\text{nlos}\nonumber\\
	\rightarrow\
	\Tr(\mathbf{R}_{kn}) &= M(\beta_{kn}^\text{los}+\beta_{kn}^\text{nlos}).\label{TR}
	\end{align}
	Here, $\beta_{kn}^\text{los}$ denotes the large-scale NLOS fading component, which we model using \cite{3GPP} 
	\begin{align}
		\beta^\text{nlos}_{kn}=\min\big\{\beta_{kn}^\text{los} \text{[dB]},&-38.3\log_{10}d\big(\bm{\zeta}^{\text{ap}}_n,\bm{\zeta}^{\text{ue}}_k\big) \nonumber\\
		& -24.9\log_{10}f-17.30\big\}\ \text{[dB]}. \label{betaNLOS}
	\end{align}		
	Additionally, the spatial correlation matrices $\{\mathbf{R}_{kn}\}$ are generated using the Gaussian local scattering model \cite[Cap. 2]{Bjornson.2017} with angular standard deviation of $10^\circ$.
	\paragraph{Transmit/Receive Parameters}
	PBs are equipped with $M=8$ antennas. Moreover, UEs transmit  $\tau_p=K$ pilot symbols with $p_k=10$ mW at each coherence time interval, while PBs' receive circuits experience an AWGN noise with $\sigma^2=-100$ dBm. UEs’ EH requirements are assumed homogeneous and equal to $1$~W (power and energy are equivalent by assuming normalized block time). 
	\begin{figure}
		\centering
		\includegraphics[width=0.46\textwidth]{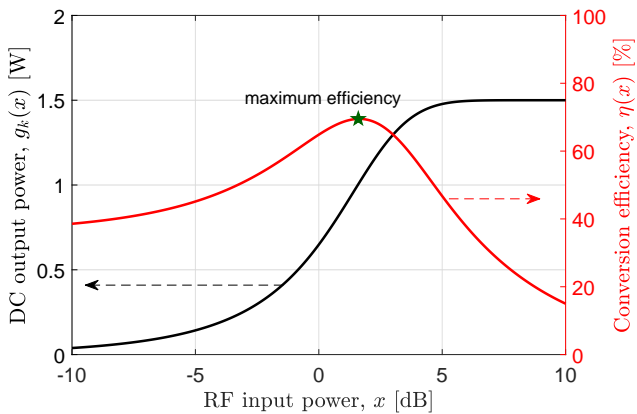}
		\caption{EH transfer function and conversion efficiency for $a=2,\ b=1$ and $\nu_k=1.5$ W.}
		\label{Fig5}
	\end{figure}
	\paragraph{EH Transfer Function}
	We consider the following EH transfer function \cite{Boshkovska.2015}: 
	\begin{align}
	g_k(x) = \nu_k \Big(\frac{1+e^{ab}}{1+e^{-a(x-b)}}-1\Big)e^{-ab},\ \forall x\ge 0,  \label{gx}
	\end{align}
	which is known to describe accurately the nonlinearity of EH circuits by properly fitting parameters $a,b\in\mathbb{R}^+$. Note that \eqref{gx} does not model the sensitivity ($\varpi_{k}$) phenomenon since $g(x)=0$ only when $x=0$. However, the  region of $x\rightarrow g^{-1}(\varpi_{k})$ in \eqref{gx} is  of little interest in practice since the EH capabilities there are very limited, especially in the scenario under discussion here where devices are power-hungry. Finally, the conversion efficiency is given by $\eta(x)=g(x)/x\in[0,1)$. Fig.~\ref{Fig5} illustrates the EH transfer function and conversion efficiency when $a=1,\ b=4$ and $\nu_k=4$ W. We use these parameter values, which do not correspond to any specific EH circuit implementation, for illustration purposes. The reasons are that EH circuits prototyping for high-power rectifying, e.g., \cite{Roberg.2012,Litchfield.2014,Wang.2015,Abbasian.2016}, are less common than for low-power implementations, which makes difficult selecting an appropriate circuit for a broad range of RF power inputs. Moreover, circuits are designed for a specific frequency range, while we will be exploring the system performance for different frequency ranges, which would require different EH circuit designs to keep approximately the same EH transfer function. Back to Fig.~\ref{Fig5}, observe the maximum conversion efficiency peak at $7.4$ dB (5.5 W) for which $\eta=59\%$. 
	Since energy requirements per transmit block are fixed, i.e., $\{\xi_k\}$ is given, the optimum system efficiency ($\min P_T$) is attained when all devices are operating at such point.
	\subsection{Unconstrained Optimization - Single UE}\label{unUE}
	Herein, we focus on the performance of the system when serving a single UE at a critical position, i.e., $\bm{\zeta}^\text{ue}=[3,3,0]^T$, which is the point that is simultaneously farthest from all the PBs. We omit the per-PB power and EMF-related constraints, thus, the MRT-based precoding specified in \eqref{v1} with $p=\delta_1/||\hat{\mathbf{h}}_1||^2$ is the optimum, thus it is used throughout this subsection.
	\begin{figure}
		\centering
		\includegraphics[width=0.46\textwidth]{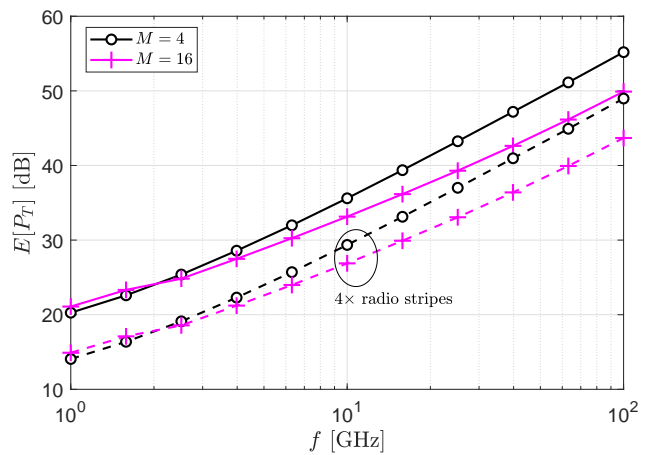}
		\caption{Average total transmit power as a function of $f$ for $M\in\{4,16\}$. The dashed curves represent the performance when the radio stripes system is four-fold vertically replicated with inter-stripes separation distance $l_0$.}
		\label{Fig6}
    \end{figure}
	\begin{figure}
		\centering
		\includegraphics[width=0.48\textwidth]{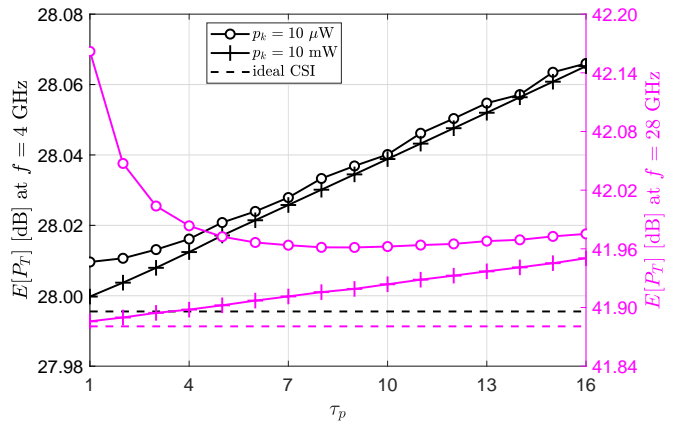} \caption{Average total transmit power as a function of the number of training pilots for $f\in\{4,28\}$ GHz.}
		\label{Fig7}
	\end{figure}

	Fig.~\ref{Fig6} shows the average transmit power requirements to serve the UE as a function of the operation frequency. For each frequency, it is assumed that the radio stripes system is composed of the maximum allowed number of PBs, which matches \eqref{N2}. Observe that as the frequency increases, the spatial transmit gains increase as well since more PBs/antennas can be incorporated to the radio stripes system. However, such gains do not compensate the path loss, which grows following a power-law with exponent $2-2.49$ according to \eqref{beta} and \eqref{betaNLOS}. Moreover, note that equipping each PB with more antennas becomes more attractive than increasing the number of PBs as the frequency increases. This is because such strategy allows obtaining a greater value of $MN$ (given the scenario geometry and inter-PB separation constraints, see \eqref{N2}), which, rather than the individual contributions from $M$ and $N$, determines the system performance when operating without per-PB power and EMF-related constraints.
	It becomes evident that additional boosting performance strategies are needed to increase the overall system efficiency, and make the technology viable in practice. For instance, by increasing $l-$fold the number of transmit antennas, the system power consumption is approximately reduced $ l-$fold as well, as illustrated in the figure for $l=4$.
		
\begin{figure}		
		\centering
		\!\!\!\!\!\!\includegraphics[width=0.46\textwidth]{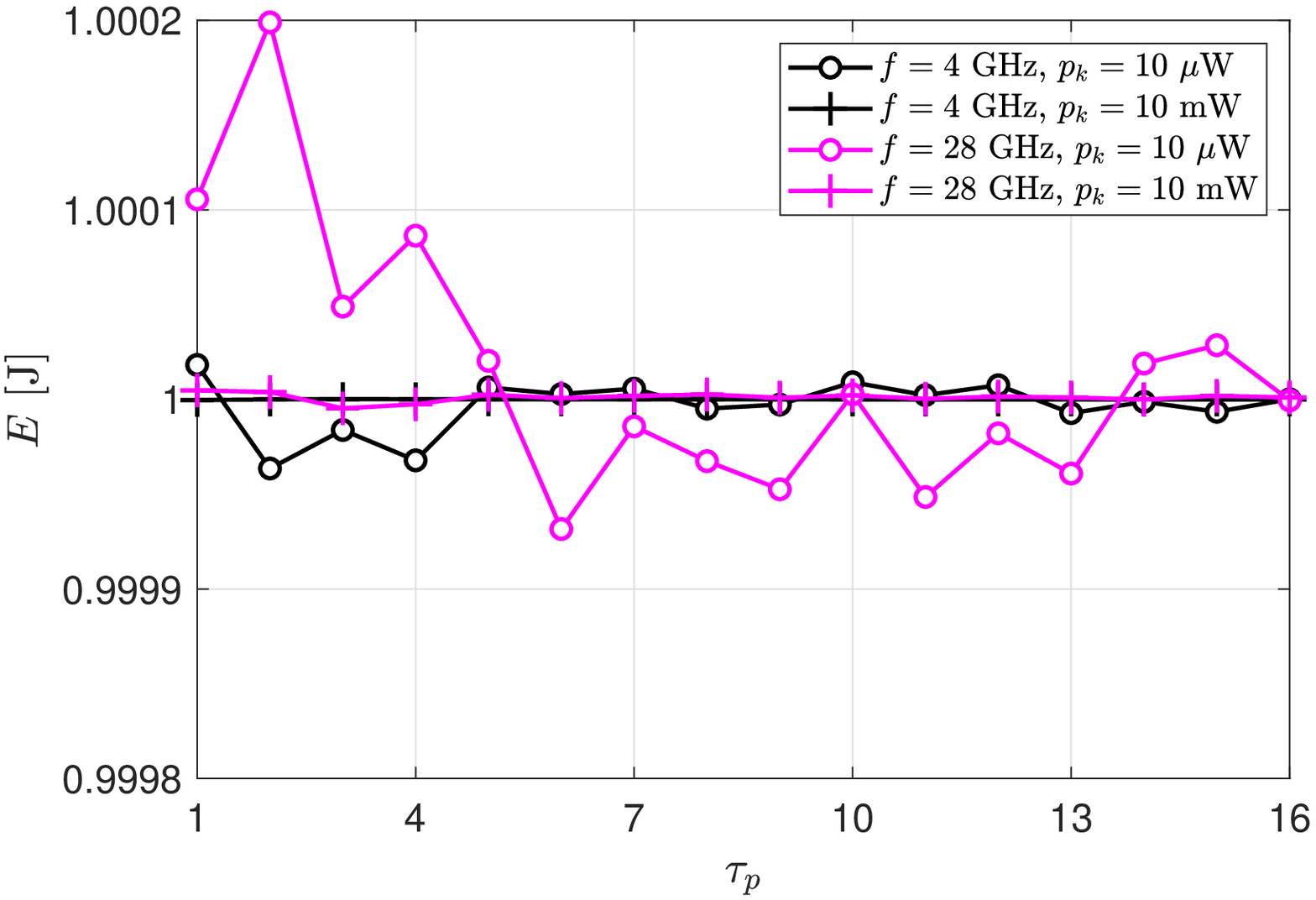}\\
		\includegraphics[width=0.48\textwidth]{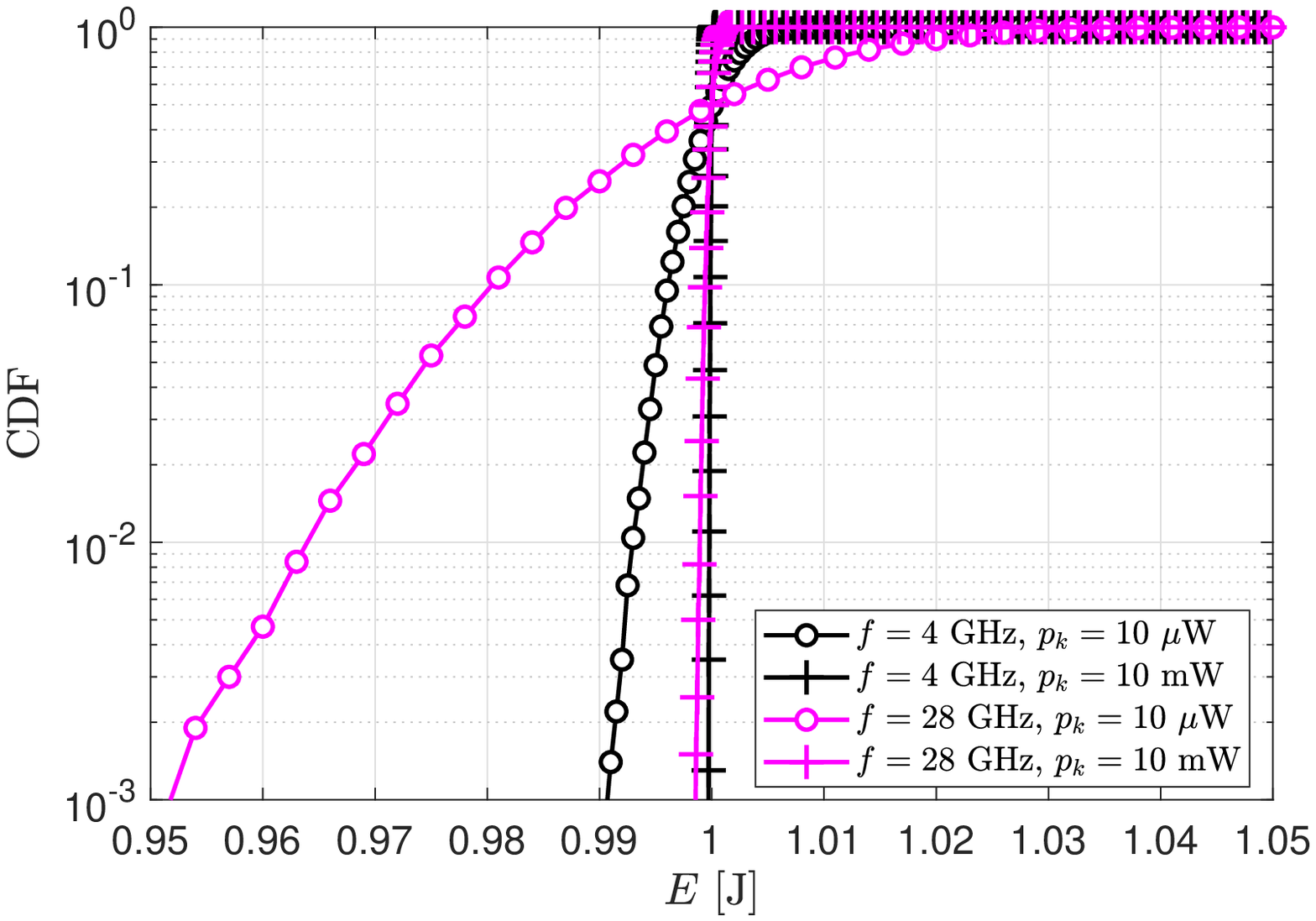}
		\caption{a) Average harvested energy as a function of $\tau_p$ (top), and b) CDF of the harvested energy for $\tau_p=1$ (bottom). We evaluate the performance for $f\in\{4,28\}$ GHz and $p_k\in\{10\ \mu\textit{W},10\ \text{mW}\}$.}
		\label{Fig8}
	\end{figure}

	Fig.~\ref{Fig7} illustrates the average transmit power consumption as a function of the duration of the CSI acquisition phase for $f\in\{4,28\}$ GHz. There is a trade-off here: the smaller $\tau_p$ is, the worse the CSI estimate becomes, however as $\tau_p$ increases, less time is left for WET; thus, both a too small or too large $\tau_p$ may affect the system efficiency. In the figure, this is most evident for $f=28$ GHz and $p_k=10\ \mu$W, for which $\tau_p^\text{opt}=9$, since   $\tau_p^\text{opt}=1$ for other configurations. Even when training power is critically reduced, e.g., to $10\ \mu$W, the average power consumption of the system does not increase significantly. Interestingly, the average harvested energy oscillates closely around the target EH requirement of $\xi_k=1$ J even when using $p_k=10\ \mu$W $@28$ GHz as illustrated in Fig.~\ref{Fig8}a. However, for rigorously meeting the EH constraint at each coherence charging block, more power resources are needed for training as evinced in Fig.~\ref{Fig8}b, where the CDF of the harvested energy is plotted for $\tau_p=1$. Note that by setting $p_k=10$ mW, the EH constraint is satisfied almost deterministically. In multi-user setups, the EH constraint is more easily satisfied since a greater $\tau_p$ is used by design. Results illustrated in Figs.~\ref{Fig7} and \ref{Fig8} support our claims at the end of Section~\ref{CSI}. In a nutshell, training is not a critical (limiting) factor for this system, which operates with orthogonal pilots, since a small amount of energy is required to attain performance similar to that of a system with ideal  CSI. Therefore, training operation configured with small number of pilots (the minimum possible, i.e., $\tau_p=K$) and small transmit power is recommended.
	
\begin{figure}
		\centering
		\includegraphics[width=0.48\textwidth]{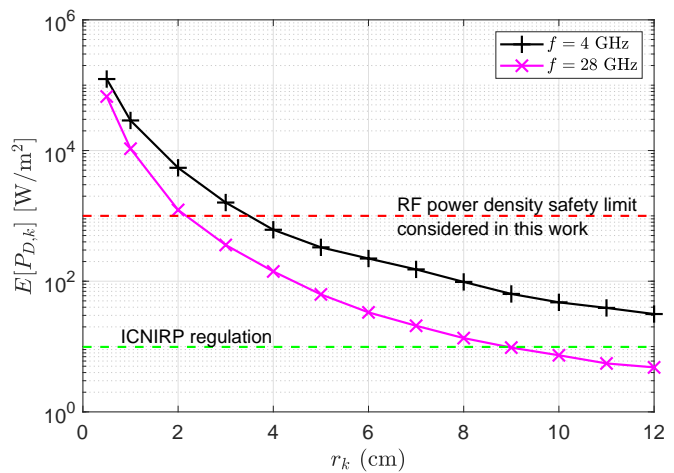}
		\caption{Average RF power density as a function of $r_k$ for $f\in\{4,28\}$ GHz.}
		\label{Fig9}
	\end{figure}

\begin{figure}
		\centering
		\!\!\!\!\!\!\! \includegraphics[width=0.46\textwidth]{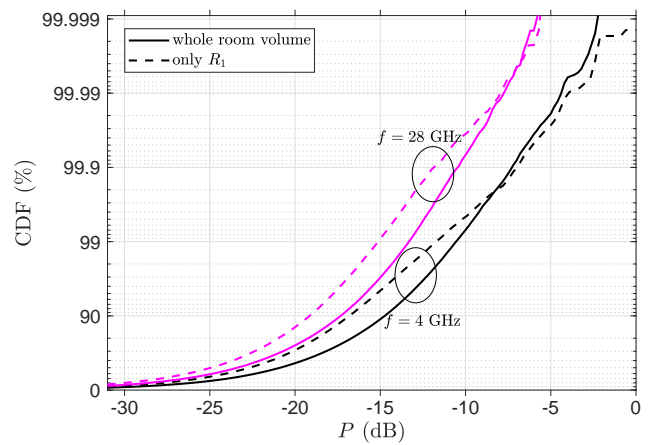}
		\caption{CDF of the RF power exposure in the room for $f\in\{4,28\}$ GHz.}
		\label{Fig10}		
	\end{figure}

	Figs.~\ref{Fig9} and~\ref{Fig10} 
	show the performance in terms of EMF exposure.  Although for higher frequency transmit power increases for the same target EH requirement at the UE to compensate the frequency-dependent losses (as shown in Figs.~\ref{Fig6} and \ref{Fig7}), the system becomes more EMF friendly. This is due to the better spatial resolution, i.e., smaller wavelength ($\lambda=1.07$ cm at $28$ GHz).
	Specifically, Fig.~\ref{Fig9} illustrates the average RF power density in the $r_k-$proximity of the UE for $f\in\{4,28\}$ GHz. As observed, the MRT-based precoding alone (without EMF-related constraints) guarantees that the RF power density at distances larger than 2 cm from the UE is below 1000~W/$\mathrm{m}^2$ at 28 GHz.\footnote{Current regulations impose much more stringent EMF radiation constraints. For instance, ICNIRP establishes the power density safety threshold at $\sim 9.9$ W$/\mathrm{m^2}$ (61 V/m) \cite{Alhasnawi.2020}, which has been depicted in Fig.~\ref{Fig9} as well. Even so, we adopt $\Theta_1=1000~$W/$\mathrm{m}^2$, above which there might be skin affectations   \cite{Tran.2017},  for our discussions in the next subsection since using smaller values would demand operating at higher frequencies, which is extremely computationally demanding and thus requires further and dedicated research. Please refer to our specific discussions on this in Section~\ref{conclusions}.} 
	Fig.~\ref{Fig10} displays the CDF of the RF power exposure in the room, i.e., $P(x,y,z)$ in \eqref{Pxyz}. Herein we observe that the chances of high RF power exposure, e.g., $\sim$1 W, at a certain spatial point in $R_1$ are significantly small.
	\subsection{Constrained Optimization}\label{multiUE}
	Herein, we consider a single UE at different locations (Fig.~\ref{Fig11} and Fig.~\ref{Fig12}), or multiple UEs (Fig.~\ref{Fig13} and Fig.~\ref{Fig14}). We consider eight different locations for the UEs as illustrated in Fig.~\ref{Fig4}. Unless stated otherwise, we set $p_\text{max}=10$ dB and the parameters related to the EMF radiation constraints as $r_k=1.5$ cm,  $\Theta_1=1000\ \text{W}/\text{m}^2$ \cite{Tran.2017}, $\varepsilon=10^{-2}$ and $\Theta_2=0.25\ $W. 
	\subsubsection{Single UE at Multiple Locations}
\begin{figure}
		\centering
		\includegraphics[width=0.46\textwidth]{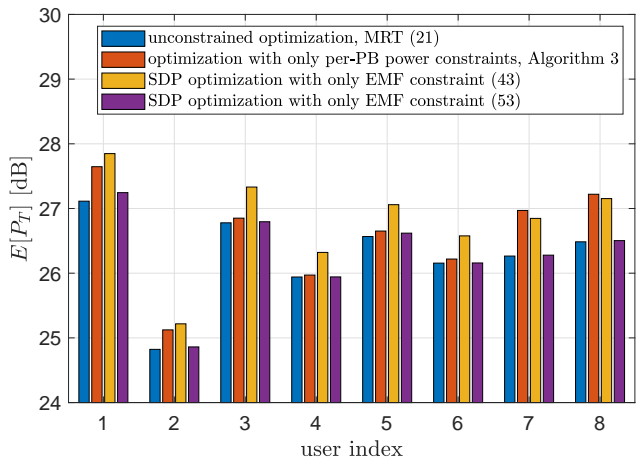}
		\caption{Average total transmit power when serving a single UE at the positions in Fig.~\ref{Fig4}. We compare the performance of the optimal i) unconstrained optimization, given by the MRT precoder in \eqref{v1}, ii) optimization with only the per-PB power constraints, given in Algorithm~\ref{alg1}, iii) SDP optimization with only the EMF constraint in \eqref{conPD}, and iv) SDP optimization with only the EMF constraint in \eqref{const2}.}
		\label{Fig11}
	\end{figure}

	Fig.~\ref{Fig11} shows the system's average transmit power when serving a single UE at different positions. Note that the UE location significantly affects the system performance. For instance,  the radio stripes system may need just $25$ dB to serve $\text{UE}_2$, while the value scales up to $27-28$ dB when serving $\text{UE}_1$. Moreover, the system incurs in a greater transmit power consumption once the per-PB power and EMF-related constraints are considered. Specifically, the per-PB power constraint affects more critically the system performance when serving $\text{UE}_1,\ \text{UE}_2,\ \text{UE}_7$ and $\text{UE}_8$, which are closer to the radio stripes  as illustrated in Fig.~\ref{Fig4}. Observe also that the EMF constraint for the proximity region of the UE is often active, thus, critically influencing the system performance by demanding an increased radio stripes' transmit power to safely diffuse the energy around a UE, while the EMF constraint for $R_1$ is not leading to a significant performance degradation of the system. 
	
\begin{figure}
		\centering
		\ \ \ \ \  \includegraphics[width=0.43\textwidth]{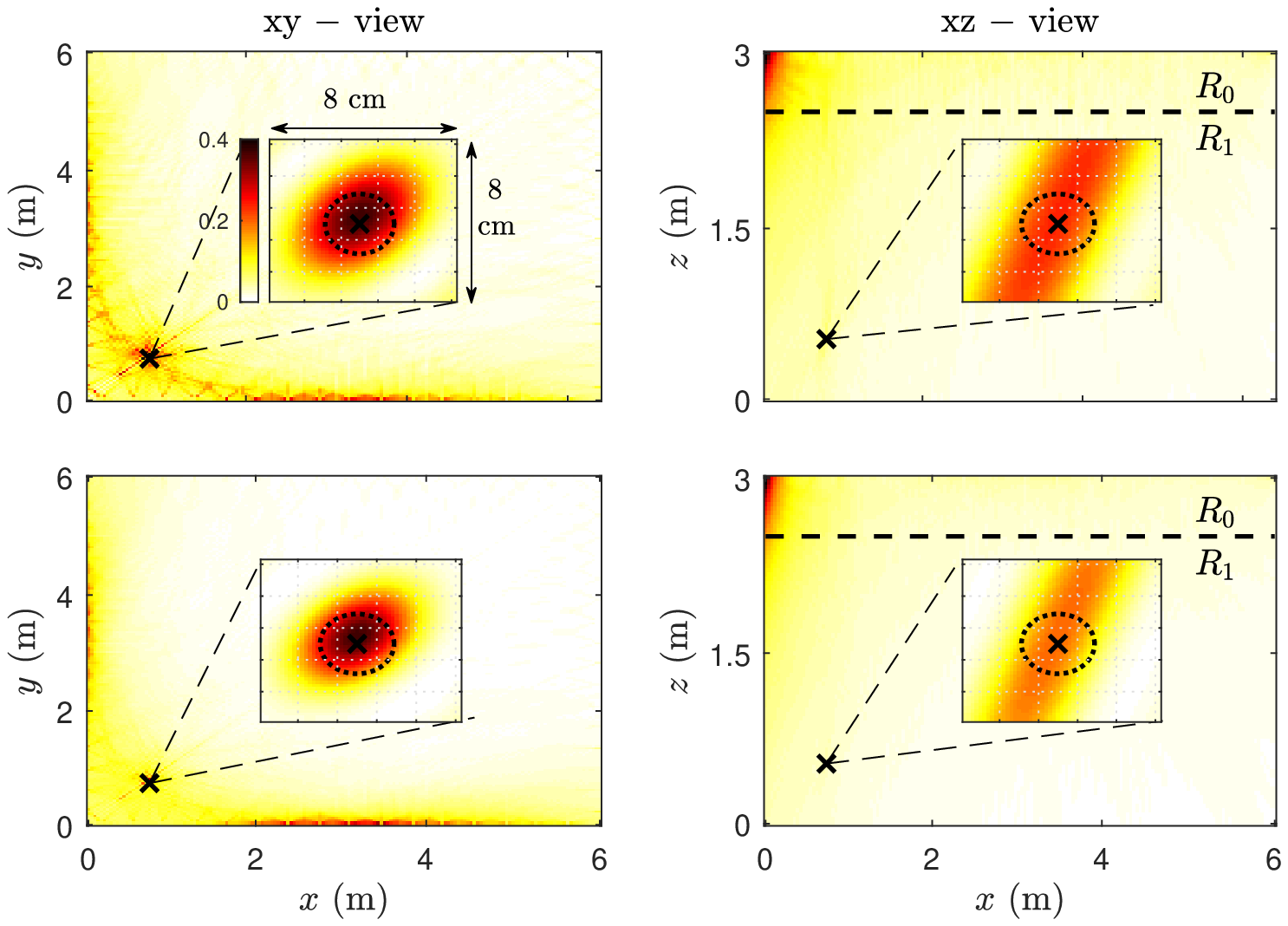}\!\!\!\!\!\!\!\!\!\!\! \includegraphics[width=0.045\textwidth]{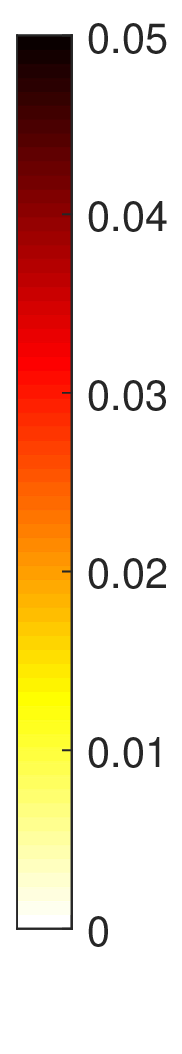}\vspace{2mm}
		\caption{Heatmap of the RF power (averaged over the missing axis, e.g., averaged over the $z-$axis when presented in the $xy-$view) in the room when powering $\text{UE}_1$. We set $p_\text{max}=17$ dB, and show the system performance when operating without (first-row plots) and with (second-row plots) EMF-related constraints. The plots corresponding to the $yz-$view are not shown as they coincide the ones for the $xz-$view  since  $x$ and $y$ coordinates of $\text{UE}_1$ match.}  
		\label{Fig12}		
	\end{figure}

	In Fig.~\ref{Fig12}, a heatmap of the RF power in the room is shown when powering only $\text{UE}_1$. We relaxed the per-PB power constraint to $p_\text{max}=17$~dB such that the main performance affectations come from the EMF-related constraints. The EMF-aware precoding somewhat homogenizes the RF radiation over the region $R_1$ compared to the precoding without EMF-related constraints. Meanwhile, the RF power density in the proximity of the UE becomes a bit less intense but more homogeneous around the center to meet the corresponding EMF constraint. 
	\subsubsection{Multiple UEs}
	In the remaining figures, we evaluate the performance in terms of average transmit power consumption and average required run time as a function of the number of UEs. Powering $k$ UEs means powering $\text{UE}_1,\cdots,\text{UE}_k$ with positions illustrated in Fig.~\ref{Fig4}. As a multi-UE system is now considered, we relax some system constraints to favor optimization feasibility, i.e., $r_k=2$~cm,  and $p_\text{max}=20$~dB. The three key optimization frameworks, say i) SDP-based precoding ($\mathbf{P2}$), ii) SCA-based precoding (Algorithm~\ref{alg1-P4}), and iii) MRT-based precoding (Algorithm~\ref{alg1-P5}), are evaluated subject to EMF constraints using CVX \cite{cvx}.
	
\begin{figure}[t!]
		\centering
		\includegraphics[width=0.46\textwidth]{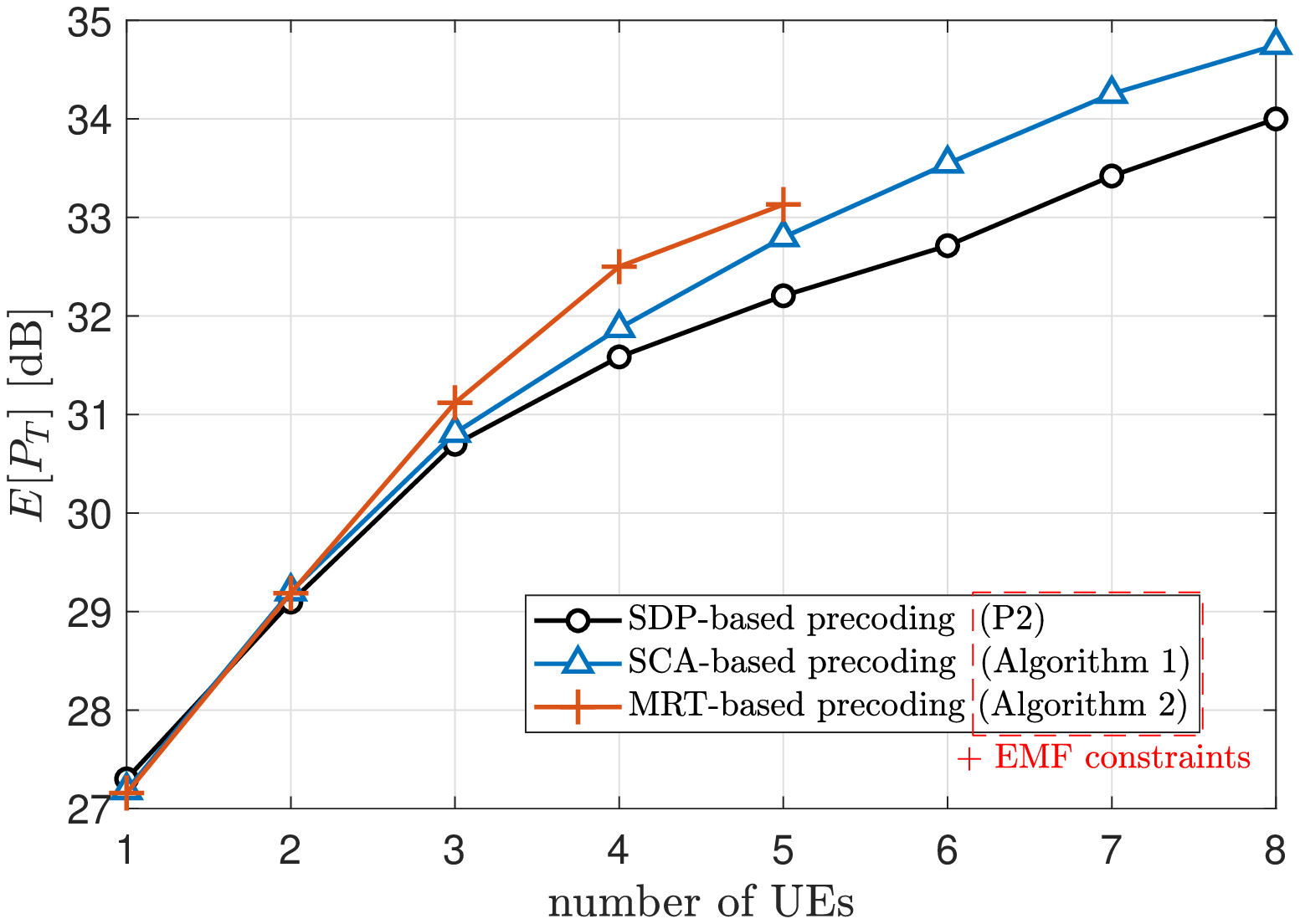}\\
		\includegraphics[width=0.46\textwidth]{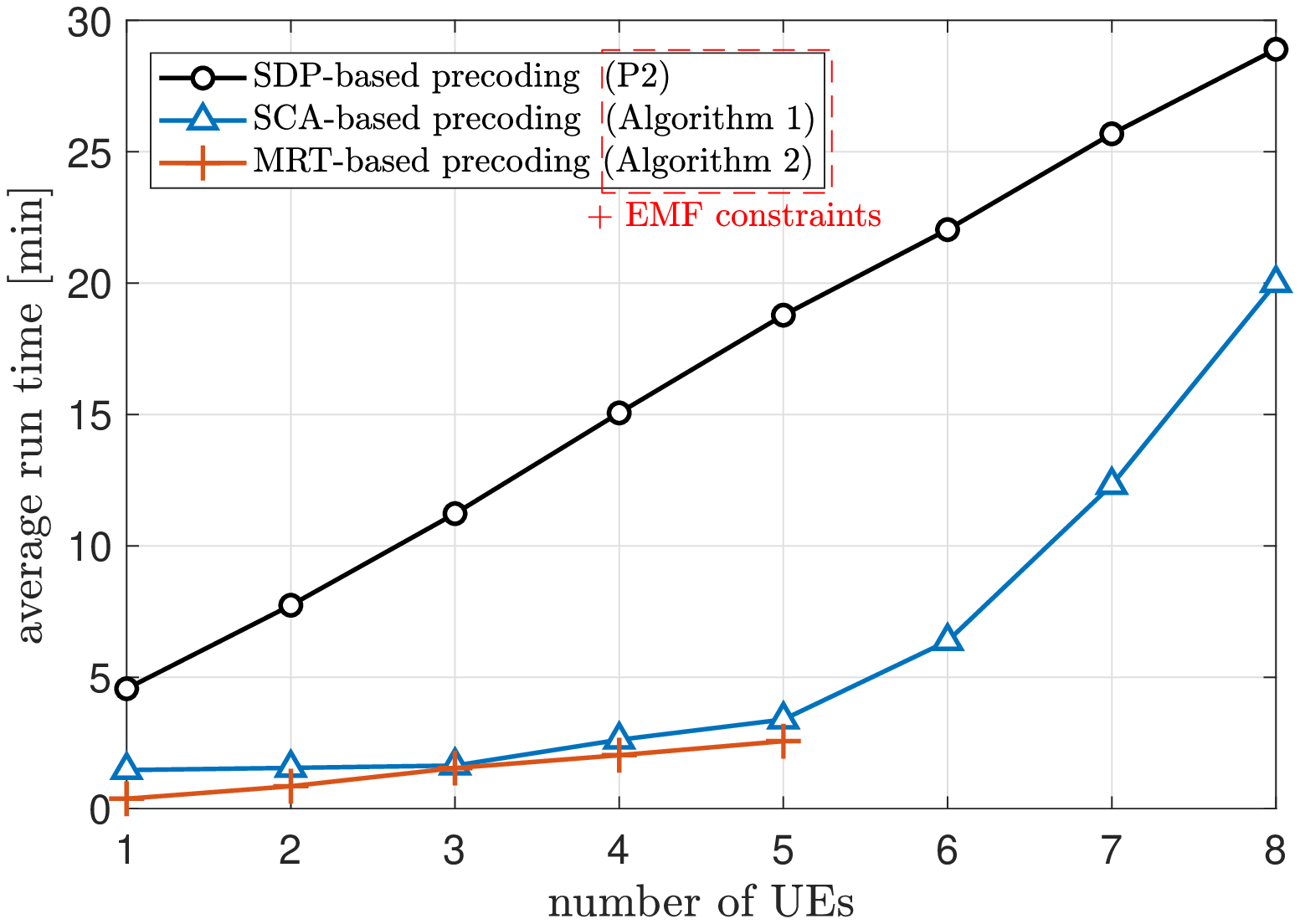}
		\caption{a) Average total transmit power (top), and b) average optimization run time (bottom) as a function of the number of UEs. 
		}
		\label{Fig13}
	\end{figure}

	Fig.~\ref{Fig13} evinces that the SDP-based framework leads to the most efficient transmit power consumption, however is extremely costly and the complexity increases linearly with the number of UEs. The SCA-based solution approximately converges to the global optimum for a small number of UEs. However, as the number of served UEs increases, locally optimum points are more often found, thus, the performance gap with respect to the global optimum solution increases. Similarly, the MRT-based framework performs extremely well when serving a small number of UEs. However, in the considered setup, serving five or more UEs is not possible when using such framework due to unfeasibility, i.e., the power domain tuning is insufficient for handling the EMF-related constraints. In such cases, the degrees of freedom related to the beamformer phase shift optimization (as used in the SDP- and SCA-based framework) are critically required.
	
	\begin{figure}[t!]
		\centering
		\!\!\!\!\includegraphics[width=0.47\textwidth]{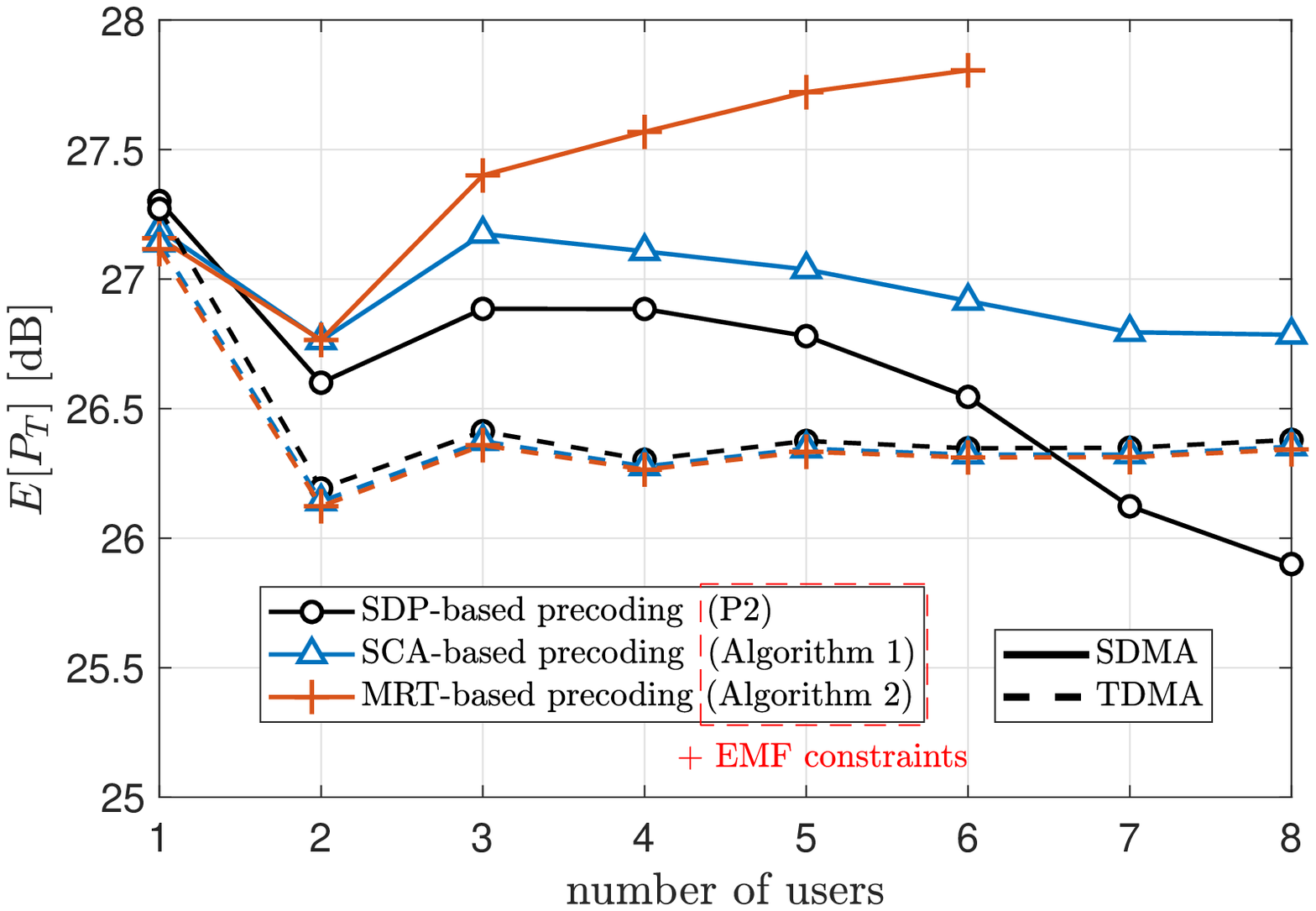}\\
		\ \includegraphics[width=0.48\textwidth]{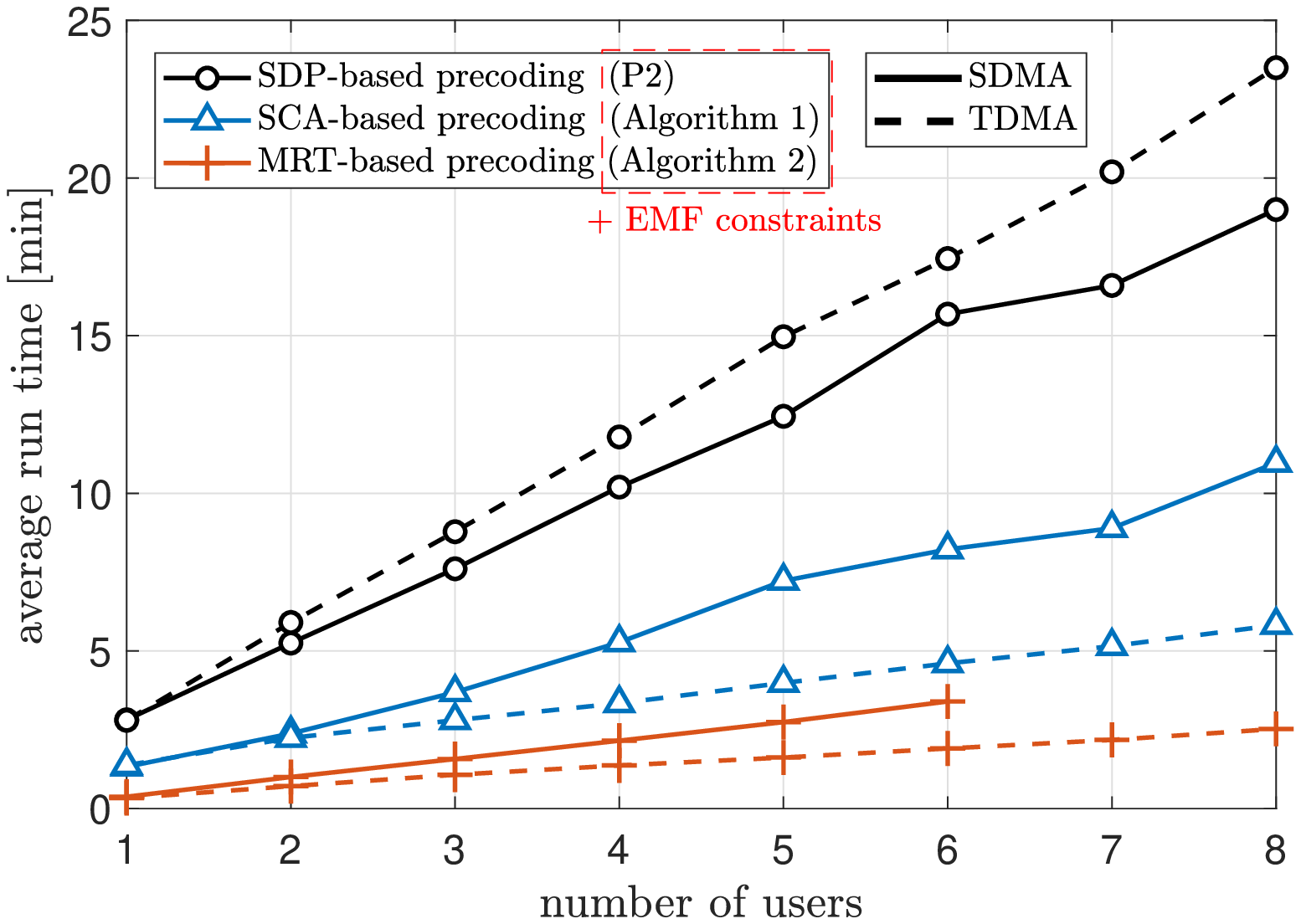}
		\caption{TDMA vs SDMA in terms of a) average total transmit power (top), and b) average optimization run time (bottom) when serving an increasing number of UEs. We set $\xi=1/k$ so the total harvested power remains at 1~W independently of the number of users.} \label{Fig14}
	\end{figure}
	In case of multi-UE scenarios, we aimed so far at powering all the UEs simultaneously by exploiting SDMA. Alternatively, the radio stripes system may exploit a TDMA technique, under which a single UE is powered at a time. However,  TDMA may be infeasible for powering any number $K$ of UEs, with a given set of energy demands $\{\xi_k\}$, as the   power required to be harvested at each $\text{UE}_k$ increases $K$ times (since charging time reduces in the same proportion), and the EH circuit may not support it, i.e., $K\xi_k> \nu_k$. Anyways, let us force feasibility by setting $\xi_k=1/K$ so the total harvested power remains at 1~W independently of the number of users, and compare TDMA vs SDMA as shown in Fig.~\ref{Fig14}. Observe that there is not functioning distinction between TDMA and SDMA for $k=1$, thus their corresponding performances match. Again, $\text{UE}_1$ is shown to be  costly to serve, especially compared to $\text{UE}_2$. Moreover,  TDMA is shown to be the most efficient in terms of average transmit power consumption when serving a relatively small number of UEs. This is mainly because, independently of the number of UEs, the EH circuit efficiency is kept high as its output is fixed at 1 W, while the output is proportionally reduced as the number of devices increases when using SDMA. Meanwhile, when the number of UEs is significantly large such lower RF incident power requirement at each UE makes easier satisfying the EMF-related constraints, thus SDMA may outperform TDMA. In terms of complexity, TDMA is shown to be more computationally costly than SDMA when using the global optimum SDP-based framework, while the opposite occurs when using SCA/MRT-based precoding. However, notice that using SDMA requires a costly single-shot optimization, while the TDMA optimization may be allowed to expand over the entire block coherence time as it is composed of a series of $k$ successive less-complex optimization processes, thus, it may be more flexible for implementation. Moreover, TDMA allows using performance and cost efficient SCA- and even MRT-based frameworks, which work well for serving a single UE at a time. In general, the performance trade-offs between SDMA and TDMA are strictly dominated by the EH non-linearity and the EMF-related constraints, and further study is needed.
	
Regarding the average optimization run time illustrated in Fig.~\ref{Fig13}b and Fig.~\ref{Fig14}b, one should only pay attention to the relative performance gap among the schemes, and not to the quantitative (absolute time) figures. This is because the absolute time performance is very sensitive to the employed solver, programming platform, and computing resources. For the discussed beamformers and scheduling mechanisms to be implementable in practice, they need to provide solution within a small portion of the channel coherence time interval. Fortunately, the channel coherence time may be usually long for the considered indoor environments, which are mostly quasi-static, thus characterized by small Doppler frequencies. To guarantee operability even in dynamic environments (small coherence time intervals), the beamformer and scheduling design must be simplified as much as possible. For this, one can rely on TDMA with a single UE served at a time and operation at very high frequency such that it might not be necessary to include EMF constraints (see discussions around Fig.~\ref{Fig9} and Fig.~\ref{Fig10}), and the fast MRT precoding design illustrated in Algorithm~\ref{alg1} can be exploited. The cost to pay is higher transmit energy consumption, thus lower system efficiency. 
\section{Conclusion and Future Works}\label{conclusions}
	We introduced an indoor mMIMO system with radio stripes for wirelessly charging power-hungry UEs. We derived optimal and sub-optimal precoders based on SDP, SCA and MRT  that aim to minimize the  radio stripes' transmit power subject to stringent EH requirements per UE, while exploiting CSI and information of the PTE of the EH circuits. Regarding this, MRT/SCA-based precoders resulted to be particularly appealing when serving a small number of UEs.
	Additionally, we proposed an  analytical framework to assess the EMF radiation exposure. We focused on the RF power density in the proximity of the UEs, and the RF power caused at a random point. These figures were incorporated as EMF-related constraints to the precoding optimization problems, thus making the WET precoding safe. Numerical results showed that  the EMF radiation exposure can be more easily controlled at higher frequencies at the cost of a higher transmit power consumption. Moreover, we verified that training is not a limiting factor for the considered indoor system as a small amount of energy is required to attain performance similar to that of a system with ideal  CSI. We discussed key trade-offs between SDMA and TDMA scheduling. TDMA is the most efficient choice in terms of average transmit power consumption when serving a relatively small number of UEs, while SDMA may be preferable  when the number of UEs is relatively large.  Some interesting research directions that we may pursue in future work are given next:
	\subsubsection{Computationally affordable methods for higher frequencies} 
	The high operation frequency (needed to enable safe WET) makes simulations extremely costly. As the operation frequency increases, the	LOS channel becomes more  oscillatory in space, thus, making rather challenging to efficiently perform  the numeric integration in \eqref{PDk}, \eqref{t1} and \eqref{t2}, even on high-tech computing servers available nowadays, taking weeks to finalize. In this work, we focused on the 4 GHz operation and thus  adopted a relatively high power density threshold, i.e., $\Theta_1=1000$ W/$\mathrm{m}^2$ (even when current regulations impose much more stringent constraints, e.g., $\sim 9.9$ W$/\mathrm{m^2}$ (61 V/m) \cite{Alhasnawi.2020}) to enable feasibility. As operation at higher frequencies may be a must, future research can focus on addressing above issues, e.g., by using principal component analysis to reduce the dimensionality of the data sets without seriously compromising accuracy \cite{Jolliffe.2016}.	
	\subsubsection{Reduce energy waste} 
	As evinced in Section~\ref{results}, the efficiency of the considered system is considerably low, usually in the range of $0.1\%-1\%$, which means that $99\%-99.9\%$ of the energy is wasted mainly due to distance/frequency/hardware-dependent losses. To make technology profitable and enable sustainability, such energy waste must be reduced as much as possible. In Fig.~\ref{Fig6}, we showed the significant performance gains from increasing the number of transmit antennas, but even greater gains may be reachable by increasing the number of receive antennas per UE. For instance, say the EH hardware is constrained on a dimension of $4$~cm~$\times~4$~cm, then, operating at $4$ GHz, $28$ GHz, $100$ GHz, allows equipping the UEs with up to 4, 71, 765 half-wavelength spaced EH antennas, respectively. 
	Since problem dimensions would increase, it may be better exploring affordable computing methods beforehand, as highlighted above.  Moreover, observe that there might be an increased energy waste when serving multiple users simultaneously, especially because the EMF-related constraints become more difficult to meet, thus intelligent scheduling mechanisms are worth investigating.
	\subsubsection{Impact of outdated/quantized CSI and radio stripes processing} 
	Channel training was shown to be not critical for the considered quasi-static indoor scenario. 
	However, the assumption of quasi-static fading is ideal per se, thus it would be interesting to evaluate a potential performance degradation due to outdated CSI, especially considering the optimization run time for each beamforming design and scheduling mechanism. Moreover, since the considered system performs centralized optimization/decisions, each PB needs to inform the estimated CSI to the CPU over limited-capacity front-haul connections. The performance losses due to quantized CSI and energy beamfomers may be significant and worth evaluating. Novel sequential \cite{Shaik.2020} or parallel \cite{Ma.2020} processing methods togheter with worst case noise designs may be attractive here.
	\subsubsection{UEs served by a reduced number of PBs}
Under ideal conditions (e.g., perfect PBs' synchronization,  fronthaul links of unlimited capacity, unlimited computation capabilities), the optimum strategy requires considering and optimizing the contribution from all PBs at every UE.
Nevertheless, the performance degradation from making UEs to be  powered by a specific set of PBs, and not all, may not be significant. In such case, the resulting trade-off between complexity and performance may tilt the scale in favor of the latter approach even under ideal conditions, while the potential is greater under non-ideal conditions. Here, proper strategies for associating PBs and UEs are needed. In LOS conditions, PBs should be associated to their closest UEs, but how may PBs to each UE? It might be strictly needed that certain UEs are dedicatedly powered by a same set of PBs to satisfy their EH demands. The probability of this increases as UEs are closer to each other, especially considering the transmit power constraints of the PBs. Finally, associating certain PBs and UEs requires using more charging signals, and also beams that are more spatially focused, which creates important challenges in terms of controlling the EMF radiation. All this must be properly addressed in a future work.
	\bibliographystyle{IEEEtran}
	\bibliography{IEEEabrv,references}
\end{document}